\documentclass[reqno,a4paper,12pt]{amsart}
\usepackage[british]{babel}
\usepackage{amsmath,amssymb,amstext,amsthm,mathrsfs,eucal,slashed,bbold}
\usepackage[utf8x]{inputenc}
\usepackage{graphicx,color}
\usepackage[a4paper,hmargin=2cm,tmargin=3cm,bmargin=3cm]{geometry}
\usepackage{array}
\usepackage{cite}
\usepackage[small]{eulervm}
\usepackage{tgpagella}
\usepackage{hyperref}
\hypersetup{%
  pdftitle   = {Supersymmetric Yang-Mills theory on conformal supergravity backgrounds in ten dimensions},
  pdfkeywords = {rigid supersymmetry, conformal superalgebra, twistor spinors, conformal Killing vectors},
  pdfauthor  = {Paul de Medeiros, José Figueroa-O'Farrill},
  pdfcreator = {\LaTeX\ with package \flqq hyperref\frqq}
}
\newcommand{\half}{\tfrac12}

\newcommand{\cL}{\mathscr{L}}

\newcommand{\fg}{\mathfrak{g}}

\newcommand{\fC}{\mathfrak{C}}

\newcommand{\fH}{\mathfrak{H}}
\newcommand{\fK}{\mathfrak{K}}

\newcommand{\fR}{\mathfrak{R}}

\newcommand{\fh}{\mathfrak{h}}
\newcommand{\fheis}{\mathfrak{heis}}
\newcommand{\fk}{\mathfrak{k}}

\newcommand{\fm}{\mathfrak{m}}

\newcommand{\fs}{\mathfrak{s}}

\newcommand{\fz}{\mathfrak{z}}

\newcommand{\fso}{\mathfrak{so}}
\newcommand{\fosp}{\mathfrak{osp}}
\newcommand{\fspin}{\mathfrak{spin}}

\newcommand{\fsl}{\mathfrak{sl}}
\newcommand{\fsp}{\mathfrak{sp}}

\newcommand{\Cl}{\mathrm{C}\ell}
\newcommand{\CW}{CW}
\newcommand{\CCl}{\mathbb{C}\ell}
\newcommand{\rd}{\mathrm{d}}

\newcommand{\Spin}{\mathrm{Spin}}

\newcommand{\1}{\mathbb{1}}

\newcommand{\RR}{\mathbb{R}}
\newcommand{\CC}{\mathbb{C}}
\newcommand{\HH}{\mathbb{H}}

\newcommand{\ZZ}{\mathbb{Z}}

\newcommand{\eD}{\mathscr{D}}

\newcommand{\eL}{\mathscr{L}}

\newcommand{\bC}{\boldsymbol{C}}
\newcommand{\bI}{\boldsymbol{I}}
\newcommand{\bP}{\boldsymbol{\Pi}}

\newcommand{\te}{\mathrm{e}}

\newcommand{\AdS}{AdS}
\newcommand{\Dirac}{{\slashed \nabla}}
\newcommand{\DiracD}{{\slashed D}}

\DeclareMathOperator{\Mat}{Mat}

\DeclareMathOperator{\diag}{diag}
\DeclareMathOperator{\vol}{vol}

\newcommand{\rn}[1]{|\!|#1|\!|}
\theoremstyle{plain}

\theoremstyle{definition}

\newcommand{\MUNCH}[1]{\relax}

\begin{document}
\title[Supersymmetric Yang--Mills and conformal
supergravity]{Supersymmetric Yang--Mills theory on conformal
  supergravity backgrounds in ten dimensions}
\author[de Medeiros]{Paul de Medeiros}
\author[Figueroa-O'Farrill]{José Figueroa-O'Farrill}
\address{Maxwell Institute and School of Mathematics, The University
  of Edinburgh, James Clerk Maxwell Building, Peter Guthrie Tait Road,
  Edinburgh EH9 3FD, United Kingdom}
\email{p.f.demedeiros@gmail.com}
\email{j.m.figueroa@ed.ac.uk}
\thanks{EMPG-15-25}
\begin{abstract}
  We consider bosonic supersymmetric backgrounds of
  ten-di\-men\-sional conformal supergravity. Up to local conformal
  isometry, we classify the maximally supersymmetric backgrounds,
  determine their conformal symmetry superalgebras and show how they
  arise as near-horizon geometries of certain half-BPS backgrounds or
  as a plane-wave limit thereof.  We then show how to define
  Yang--Mills theory with rigid supersymmetry on any supersymmetric
  conformal supergravity background and, in particular, on the
  maximally supersymmetric backgrounds.  We conclude by commenting on
  a striking resemblance between the supersymmetric backgrounds of
  ten-dimensional conformal supergravity and those of
  eleven-dimensional Poincaré supergravity.
\end{abstract}
\maketitle
\vspace*{-1.6cm}
\tableofcontents

\section{Introduction}
\label{sec:introduction}

Supersymmetry multiplets in ten-dimensional spacetime not only
underpin the five critical string theories (and their respective
low-energy supergravity limits)  but also encode the intricate
structure of extended supersymmetry in many interesting quantum field
theories in lower dimensions. For example, the Yang--Mills
supermultiplet in ten dimensions elegantly captures the structure of
extended supersymmetry and R-symmetry for gauge couplings in lower
dimensions. Of course, in dimensions greater than four, even
supersymmetric quantum field theories are not expected to be
renormalisable without some kind of non-perturbative UV completion
(indeed, this is precisely what string theory aims to
provide). Without this completion, they should merely be regarded as
low-energy effective field theories.

In addition to the more familiar (gauged) type I,
(Romans\footnote{This epithet is added when the zero-form RR flux in
  the type IIA gravity supermultiplet is non-zero.} ) type IIA and
type IIB Poincaré gravity supermultiplets \cite{Chapline:1982ww,
  CampbellWestIIA, GianiPerniciIIA, HuqNamazieIIA, Romans-Massive,
  SchwarzWestIIB, SchwarzIIB, HoweWestIIB} associated with critical
string theory, there is also a conformal gravity supermultiplet in ten
dimensions \cite{Bergshoeff:1982az}. This conformal gravity
supermultiplet can be gauged and the coupling described in
\cite{Bergshoeff:1982az} to a Yang--Mills supermultiplet in ten
dimensions is reminiscent of the analogous Chapline--Manton
\cite{Chapline:1982ww} coupling for type I supergravity. Unlike the
Poincaré supergravity theories in ten dimensions though, this
conformal supergravity theory is manifestly off-shell and must be
supplemented with some differential constraints in order to render it
local. As a supergravity theory, it is therefore somewhat exotic but
admits a consistent truncation to type I supergravity and reduces
correctly to known extended conformal supergravity theories in both
four and five dimensions. There is also a little conceptual deviation
from the unextended conformal gravity supermultiplets in lower
dimensions which result from gauging one of the conformal
superalgebras on Nahm's list \cite{Nahm:1977tg}. Of course, this is
not surprising since there are no conformal superalgebras of the
conventional type above dimension six.\footnote{By this we mean that
  there exists no real Lie superalgebra obeying the axioms of
  \cite{Nahm:1977tg} whose even part is of the form
  $\fso(s+1,t+1) \oplus \fR$, for any real Lie algebra $\fR$, if
  $s+t>6$.  Similarly, but with different hypotheses, for $n>6$, the
  maximal transitive prolongation of the $\ZZ$-graded complex Lie
  superalgebra $\fh=\fh_{-1} \oplus \fh_{-2}$, with $\fh_{-1}$ a (not
  necessary irreducible) spinor module of $\fso_n(\CC)$ and
  $\fh_{-2} = \CC^n$, the vector representation, has no pieces in
  positive degree \cite{MR3255456}.}  There do exist more general
notions of a conformal superalgebra where the conformal algebra is
contained in a less obvious manner. In particular, it was shown in
\cite{vanHolten:1982mx} that the Lie superalgebra $\fosp(1|32)$ can be
thought of as a conformal superalgebra for $\RR^{9,1}$ with respect to
a particular $\fso(10,2) < \fosp(1|32)$. Alas, it remains unclear
though whether conformal supergravity in ten dimensions is somehow
related to gauging this $\fosp(1|32)$.

There is a vast literature on the classification of supersymmetric
solutions of supergravity theories in diverse dimensions: that is to
say, backgrounds which preserve some amount of rigid supersymmetry and
solve the supergravity field equations. Indeed, at least for Poincaré
supergravities, it is often the case that the preservation of a
sufficient amount of rigid supersymmetry will guarantee that all of
the supergravity field equations are satisfied. This typically comes
from the so-called integrability conditions which result from
iterating the `Killing spinor' equations imposed by the
preservation of supersymmetry.

In recent years, there has been mounting interest in the somewhat
broader task of classifying supersymmetric backgrounds of conformal
and Poincaré supergravity theories (which need not necessarily solve
the field equations, only the integrability conditions). This is
motivated primarily by a renewed curiosity in the general structure of
quantum field theories with rigid supersymmetry in curved space
\cite{Blau:2000xg,Pestun:2007rz, Kapustin:2009kz, Jafferis:2010un,
  Jafferis:2011zi, Festuccia:2011ws, Jia:2011hw,
  Samtleben:2012gy, Klare:2012gn, Dumitrescu:2012ha, Cassani:2012ri,
  Liu:2012bi,Ito:2012hs,
  deMedeiros:2012sb,Fujitsuka:2012wg,Dumitrescu:2012at,
  Kehagias:2012fh, Closset:2012ru, Martelli:2012sz, Samtleben:2012ua,
  Kuzenko:2012vd, Hristov:2013spa, deMedeiros:2013pps,
  deMedeiros:2013mca, Martelli:2013aqa, Cassani:2013dba,
  Alday:2013lba, Ito:2013eva, Kuzenko:2013gva, Klare:2013dka,
  Pan:2013uoa, Closset:2013vra, Nosaka:2013cpa, Closset:2013sxa,
  Figueroa-O'Farrill:2013tpa, Deger:2013yla, Kuzenko:2013uya,
  Cassani:2014zwa, DiPietro:2014moa, Kuzenko:2014mva, Anous:2014lia,
  Imamura:2014ima, Farquet:2014kma, Alday:2014rxa, Assel:2014paa,
  Alday:2014bta, Kuzenko:2014eqa, Farquet:2014bda, deMedeiros:2014hla,
  Pan:2014bwa, Butter:2014gha, Kuzenko:2014yia, Closset:2014uda,
  DiPietro:2014bca, Nishioka:2014zpa, LischewskiCSA, Gomis:2014woa,
  Yoshida:2014ssa, Assel:2014tba, Butter:2014xxa, Bawane:2014uka,
  Sinamuli:2014lma, Rodriguez-Gomez:2014eza, Knodel:2014xea,
  Cabo-Bizet:2014nia, Lorenzen:2014pna, Minahan:2015jta,
  Assel:2015nca, Alday:2015lta, Pan:2015nba, Benini:2015noa,
  Pini:2015xha, Honda:2015yha, Guarino:2015jca, Kuzenko:2015lca, Butter:2015tra,
  Alday:2015jsa, Kawano:2015ssa, Fluder:2015eoa, Cassani:2015upa,
  Qiu:2015rwp, Martelli:2015kuk, Aharony:2015hix, Minahan:2015any,
  Lischewski:2015jna}, for which supersymmetric localisation has
substantiated many important exact results and novel holographic
applications \cite{Jafferis:2011zi, Klare:2012gn,
  Cassani:2012ri, Martelli:2012sz, Hristov:2013spa, Martelli:2013aqa,
  Cassani:2014zwa, Farquet:2014kma, Alday:2014rxa, Alday:2014bta,
  Farquet:2014bda, Yoshida:2014ssa, Alday:2015lta, Guarino:2015jca, Alday:2015jsa,
  Fluder:2015eoa, Cassani:2015upa}. The general strategy for obtaining
non-trivial background geometries which support rigid supersymmetry
builds on the pioneering work of Festuccia and Seiberg in four
dimensions \cite{Festuccia:2011ws}. Given a rigid supermultiplet in
flat space, it is often possible to promote it to a local
supermultiplet in curved space via an appropriate supergravity
coupling. For example, such a coupling can be induced holographically
in a superconformal field theory in flat space that is dual to a
string theory in an asymptotically anti-de Sitter background. A
judicious choice of decoupling limit (in which the Planck mass becomes
infinite) typically ensures that the dynamics of the gravity
supermultiplet are effectively frozen out, leaving only the fixed
bosonic supergravity fields as data encoding the geometry of the
rigidly supersymmetric curved background.

The aim of this paper is to explore various aspects of bosonic
supersymmetric backgrounds of conformal supergravity in ten dimensions
and elucidate the structure of the rigid Yang--Mills supermultiplet on
these backgrounds. In particular, we will classify the maximally
supersymmetric conformal supergravity backgrounds, compute their
associated conformal symmetry superalgebras and show how they are
related to each other via certain algebraic limits. We will also show
how to ascribe to any conformal supergravity background a conformal
Killing superalgebra that is generated by its Killing spinors. Paying
close attention to the non-trivial Weyl symmetry which acts within
this class of conformal supergravity backgrounds, we will see how to
recover the subclass of type I supergravity backgrounds and how
certain Weyl-transformed versions of the half-BPS string and
five-brane backgrounds of type I supergravity recover, in the
near-horizon limit, the maximally supersymmetric conformal
supergravity backgrounds of Freund--Rubin type. We will then describe
the rigid supersymmetry transformations and invariant lagrangian for
the Yang--Mills supermultiplet on any bosonic supersymmetric conformal
supergravity background. This will be done both on-shell and in the
partially off-shell formalism of \cite{Berkovits:1993zz, Evans:1994cb,
  Baulieu:2007ew}. We conclude with a curious observation that several
highly supersymmetric conformal supergravity backgrounds in ten
dimensions can be embedded in solutions of eleven-dimensional Poincaré
supergravity which preserve twice as much supersymmetry.

This paper is organised as follows. In
Section~\ref{sec:conf-supergr-backgr} we discuss supersymmetric
backgrounds of ten-dimensional conformal supergravity. In
Sections~\ref{sec:10dCSGoffshell} and
\ref{sec:10dCSGoffshellSUSYbackgrounds} we discuss the conformal
gravity supermultiplet, the Killing spinor equation and its
integrability condition. In Section~\ref{sec:10dCSKillingSuperalgebra}
we define the notion of a conformal symmetry superalgebra and show
that every supersymmetric conformal supergravity background admits a
conformal Killing superalgebra, which we define to be the ideal of a
conformal symmetry superalgebra that is generated by the Killing
spinors of the background. In Section~\ref{sec:10dCSGoffshellMaxSUSY}
we classify those conformal supergravity backgrounds preserving
maximal supersymmetry. The results mimic those of eleven-dimensional
supergravity: besides the (conformally) flat background, we have a
pair of Freund--Rubin families and their plane-wave limit. In
Sections~\ref{sec:10dCSMaximalSUSYSuperalgebra},
\ref{sec:ConformalSuperalgebrasAdSS} and
\ref{sec:ConformalSuperalgebrasPW} we work out the conformal symmetry
superalgebras of these backgrounds and show in
Section~\ref{sec:Contractions} that the Killing superalgebra of the
plane-wave limit arises as an Inönü--Wigner contraction of the Killing
superalgebra of the Freund--Rubin backgrounds. In
Section~\ref{sec:MaximalSA} we comment on the maximal superalgebra of
the maximally supersymmetric backgrounds, showing that it is
isomorphic to $\fosp(1|16)$ for the Freund--Rubin backgrounds and
non-existent for their plane-wave limit. In
Section~\ref{sec:10dCSGoffshellHalfBPSSUSY} we discuss some of the
half-BPS backgrounds of conformal supergravity and show how the
Freund--Rubin maximally supersymmetric backgrounds arise as
near-horizon geometries. In Section~\ref{sec:10dSYMonshell} we
introduce the on-shell Yang--Mills supermultiplet and write down a
supersymmetric lagrangian on any supersymmetric conformal supergravity
background. We then describe a partially off-shell formulation of
supersymmetric Yang--Mills theory on any such background. Finally, in
Section~\ref{sec:11dLift} we explore a possible relation between
ten-dimensional conformal supergravity and eleven-dimensional Poincaré
supergravity suggested by the resemblance between their maximally
supersymmetric backgrounds and some of the half-BPS backgrounds of
both theories. Appendix~\ref{sec:10dCliffordSpinors} contains our
Clifford algebra conventions.


\section{Conformal supergravity backgrounds}
\label{sec:conf-supergr-backgr}

\subsection{Conformal gravity supermultiplet}
\label{sec:10dCSGoffshell} 

The off-shell conformal gravity supermultiplet in ten dimensions was
constructed in \cite{Bergshoeff:1982az}. The bosonic sector contains
a metric $g_{\mu\nu}$, a six-form gauge potential $C_{\mu_1 \dots \mu_6}$
and an auxiliary scalar $\phi$. The fermionic sector contains a
gravitino $\psi_\mu$ and an auxiliary spinor $\chi$. Both $\psi_\mu$
and $\chi$ are Majorana--Weyl spinor-valued, with opposite
chiralities.\footnote{Our spinor conventions are contained in
  Appendix~\ref{sec:10dCliffordSpinors}.}  The bosonic fields
$g_{\mu\nu}$ and $C_{\mu_1 \dots\mu_6}$ contribute $44{+}84$ off-shell
degrees of freedom, matching the $8{\times}16$ off-shell degrees of
freedom from the fermionic field $\psi_\mu$. The fields
$(g_{\mu\nu},C_{\mu_1 \dots\mu_6},\phi,\psi_\mu,\chi)$ are assigned Weyl
weights $(2,0,w,\half,-\half)$.

The supersymmetry variations for this theory can be found in equation
(3.34) of \cite{Bergshoeff:1982az} and must be supplemented with the
constraint defined in their equation (3.35). Their `Q' and `S'
supersymmetry parameters are described by a pair of Majorana--Weyl
spinors $\epsilon$ and $\eta$ with opposite chiralities: for
definiteness, we shall take $\epsilon$ to have positive chirality,
i.e., $\Gamma \epsilon = \epsilon$.  A bosonic supersymmetric
background of this theory follows by solving the equations obtained by
setting to zero the combined `Q' and `S' supersymmetry variation of
$\psi_\mu$ and $\chi$, evaluated at $\psi_\mu = 0$ and $\chi =0$.

On a ten-dimensional lorentzian manifold $(M,g)$ equipped with
Levi-Civita connection $\nabla$, the equations which follow from
this procedure are 
\begin{equation}
  \label{eq:10dCKS}
  \begin{split}
    \nabla_\mu \epsilon + \tfrac{1}{4} \phi^{6/w} ( \Gamma_\mu
    K + 2 K \Gamma_\mu ) \epsilon &= \Gamma_\mu \eta \\
    \tfrac{1}{2w} \phi^{-1} ( \Dirac \phi ) \epsilon
    +\tfrac{1}{12} \phi^{6/w} K \epsilon &= \eta~,
  \end{split}
\end{equation}
where $K =\rd C$. The constraint in equation (3.36) of
\cite{Bergshoeff:1982az} follows as an integrability condition from
\eqref{eq:10dCKS}. Notice that the second equation in
\eqref{eq:10dCKS}, derived from the supersymmetry variation of $\chi$,
is simply a definition of $\eta$ in terms of the other background
data. Substituting this definition into the first equation  in
\eqref{eq:10dCKS} thus yields the defining condition for a bosonic
supersymmetric background.

\subsection{Supersymmetric backgrounds}
\label{sec:10dCSGoffshellSUSYbackgrounds} 

Let us now define a more convenient set of background fields to work
with:
\begin{equation}
  \label{eq:10dCSGredef}
  \Phi := \tfrac{3}{w} \ln \phi \qquad\text{and}\qquad H := 4 \,
  \phi^{6/w} {\star K}~,
\end{equation}
and write $G = \rd \Phi$. The three-form $H$ obeys $\rd (\te^{-2\Phi}
{\star H} )=0$ since $K$ is a closed seven-form. In terms of this data, the
defining condition \eqref{eq:10dCKS} for a bosonic supersymmetric
background $(M,g,G,H)$ of conformal supergravity in ten dimensions becomes
\begin{equation}
  \label{eq:10dCSGvacua}
  \nabla_\mu \epsilon = \tfrac{1}{6} \Gamma_\mu G \epsilon +
  \tfrac{1}{24} \Gamma_\mu H \epsilon + \tfrac{1}{8} H \Gamma_\mu \epsilon~.
\end{equation}

Under a Weyl transformation $g_{\mu\nu} \mapsto \Omega^2 g_{\mu\nu}$,
for some positive function $\Omega$, it follows that $\Gamma_{\mu}
\mapsto \Omega \Gamma_{\mu}$ and $\epsilon \mapsto \sqrt{\Omega}
\epsilon$. The condition \eqref{eq:10dCSGvacua} is therefore preserved
under any such transformation provided $H \mapsto \Omega^2 H$ and
$\Phi \mapsto \Phi + 3 \ln \Omega$.  Consequently, performing this
transformation with $\Omega = \te^{-\Phi/3}$ allows one to
fix $G=0$ in equation~\eqref{eq:10dCSGvacua} with $H$ coclosed:
\begin{equation}
  \label{eq:10dCSGvacuaGzero}
  \nabla_\mu \epsilon = \tfrac{1}{24} \Gamma_\mu H \epsilon +
  \tfrac{1}{8} H \Gamma_\mu \epsilon \qquad\text{and}\qquad \rd {\star H}
  =0~.
\end{equation}

The condition \eqref{eq:10dCSGvacua} implies that the `Dirac current'
one-form $\xi_\mu = {\overline \epsilon} \Gamma_\mu \epsilon$ and the
self-dual five-form $\zeta_{\mu\nu\rho\sigma\tau} = {\overline
  \epsilon} \Gamma_{\mu\nu\rho\sigma\tau} \epsilon$ obey
\begin{equation}
  \label{eq:10dCSGvacuaCKV}
  \nabla_\mu \xi_\nu = \tfrac{1}{3} g_{\mu\nu} G_\xi + \tfrac{1}{3}
  \left( H_{\mu\nu\rho} \xi^\rho + 2 G_{[\mu} \xi_{\nu]} +
    \tfrac{1}{12} \zeta_{\mu\nu\rho\sigma\tau} H^{\rho\sigma\tau}
  \right)~,
\end{equation}
and
\begin{equation}
  \label{eq:10dCSGvacuaSD5form}
  \nabla^\tau \zeta_{\mu\nu\rho\sigma\tau} = 2
  \zeta_{\mu\nu\rho\sigma\tau} G^\tau -4 H_{[\mu\nu\rho}
  \xi_{\sigma]}~.
\end{equation} 

Taking the $(\mu\nu)$ symmetric part of \eqref{eq:10dCSGvacuaCKV}
implies $\cL_\xi g = -2\sigma_\xi g$ with
\begin{equation}
  \label{eq:10dCSGvacuaSigmaxi}
  \sigma_\xi = - \tfrac{1}{10} \nabla_\mu \xi^\mu = -\tfrac{1}{3}
  \eL_\xi \Phi~,
\end{equation} 
which shows that $\xi$ is a conformal Killing vector. Furthermore,
acting with $\nabla^\sigma$ on \eqref{eq:10dCSGvacuaSD5form} and using
closure of $\te^{-2\Phi} {\star H}$ and $G$  together with
\eqref{eq:10dCSGvacuaCKV} and \eqref{eq:10dCSGvacuaSD5form} on the
right hand side implies
\begin{equation}
  \label{eq:10dCSGvacuaLiexiH}
  \eL_\xi H = -2 \sigma_\xi H~.
\end{equation} 

If $H$ is closed then solutions of \eqref{eq:10dCSGvacua} with
$G \epsilon = \half H \epsilon$ describe bosonic
supersymmetric backgrounds of type I supergravity in ten dimensions.
In that case, \eqref{eq:10dCSGvacua} reduces to $\nabla_\mu \epsilon =
\tfrac{1}{8} H_{\mu\nu\rho} \Gamma^{\nu\rho} \epsilon$, $\xi$ is a
Killing vector and $\iota_\xi H$ is closed. Clearly any such
background is a special case of \eqref{eq:10dCSGvacua} and so one can
always perform a Weyl transformation to obtain a solution of
\eqref{eq:10dCSGvacuaGzero}.  However, if the original background had
$G \neq 0$ then the new supersymmetric background of conformal
supergravity solving \eqref{eq:10dCSGvacuaGzero} will no longer be a
supersymmetric background of type I supergravity since the required
Weyl transformation does not preserve the defining conditions $\rd
H=0$ and $G \epsilon = \half H \epsilon$.

Evaluating $[\nabla_\mu,\nabla_\nu]\epsilon$ implies the integrability condition
\begin{multline}
  \label{eq:10dCSGInt}
  \tfrac{1}{4} \left( R_{\mu\nu\rho\sigma} -\tfrac{2}{3} \nabla_{[\mu}
    H_{\nu]\rho\sigma} - \tfrac{1}{6} H_{\mu\rho\alpha}
    H_{\nu\sigma}{}^{\alpha} -\tfrac{4}{9} H_{\mu\nu\rho} G_\sigma
  \right) \Gamma^{\rho\sigma} \epsilon +\tfrac{1}{18} \left(
    \tfrac{1}{24} H_{\alpha\beta\gamma} H^{\alpha\beta\gamma}  -
    G_\alpha G^\alpha \right) \Gamma_{\mu\nu} \epsilon \\
  -\tfrac{1}{3} \left( \nabla_{[\mu} G^\rho + \tfrac{1}{3} G_{[\mu}
    G^\rho - \tfrac{1}{24} H^{\rho}{}_{\alpha\beta}
    H_{[\mu}{}^{\alpha\beta} - \tfrac{1}{3} G_\alpha
    H_{[\mu}{}^{\rho\alpha} \right) \Gamma_{\nu]\rho} \epsilon +
  \tfrac{1}{36} \left( H^{\rho\sigma\alpha} G_\alpha \right) 
  \Gamma_{\mu\nu\rho\sigma} \epsilon \\
  +\tfrac{1}{108} \left( H_{\mu\nu\alpha} H_{\beta\gamma\delta} +
    \tfrac{1}{16} H_{[\mu}{}^{\rho\sigma} {\star
    H}_{\nu]\rho\sigma\alpha\beta\gamma\delta} + \tfrac{1}{64}
    \varepsilon_{\mu\nu\alpha\beta\gamma\delta}{}^{\rho\sigma\theta\phi}
    H_{\rho\sigma\epsilon} H_{\theta\phi}{}^\epsilon \right)
  \Gamma^{\alpha\beta\gamma\delta} \epsilon \\
  +\tfrac{1}{72} \left( \nabla_{\mu} H_{\alpha\beta\gamma} +
    H_{\mu\alpha\beta} G_\gamma + \tfrac{1}{3} H_{\alpha\beta\gamma}
    G_\mu - H_{\mu\alpha}{}^\rho H_{\beta\gamma\rho} \right)
  \Gamma_{\nu}{}^{\alpha\beta\gamma} \epsilon \\
  -\tfrac{1}{72} \left( \nabla_{\nu} H_{\alpha\beta\gamma} +
    H_{\nu\alpha\beta} G_\gamma + \tfrac{1}{3} H_{\alpha\beta\gamma}
    G_\nu - H_{\nu\alpha}{}^\rho H_{\beta\gamma\rho} \right)
  \Gamma_{\mu}{}^{\alpha\beta\gamma} \epsilon =0~,
\end{multline}
for every $\epsilon$ solving \eqref{eq:10dCSGvacua}.  The geometric
meaning of this equation is the following.
Equation~\eqref{eq:10dCSGvacua} defines a connection $\eD$ on the
spinor bundle by declaring that a spinor $\epsilon$ is $\eD$-parallel
if and only if it satisfies equation~\eqref{eq:10dCSGvacua}.  Then
equation~\eqref{eq:10dCSGInt} is simply the statement that
$\eD$-parallel spinors are invariant under the holonomy algebra of
$\eD$ and, in particular, are annihilated by the curvature of $\eD$.

\subsection{Conformal symmetry superalgebras}
\label{sec:10dCSKillingSuperalgebra}

Let $(M,g,G,H)$ be a bosonic supersymmetric background of conformal
supergravity in ten dimensions. Let $\fC(M,g)$ denote the Lie algebra
of conformal Killing vectors on the ten-dimensional lorentzian
manifold $(M,g)$. The Lie subalgebra of homothetic conformal Killing
vectors will be written $\fH(M,g) < \fC(M,g)$ which contains as an
ideal the Lie algebra of Killing vectors $\fK(M,g) \lhd \fH(M,g)$.

Now let us ascribe to $(M,g,G,H)$ a $\ZZ_2$-graded vector space
$\fs = \fs_{\bar 0} \oplus \fs_{\bar 1}$, with even part
$\fs_{\bar 0} \subset \fC(M,g)$ and odd part $\fs_{\bar 1} = \ker \eD$
spanned by solutions $\epsilon$ of \eqref{eq:10dCSGvacua}. We would
like to equip $\fs$ with the structure of a Lie superalgebra. The
first step is to define a bracket on $\fs$, i.e., a skewsymmetric (in
the graded sense) bilinear map
$[-,-] : \fs \times \fs \rightarrow \fs$ such that
\begin{equation}
  \label{eq:10dCSLieBracket}
  [\fs_{\bar 0},\fs_{\bar 0}] \subset \fs_{\bar 0} \; , \qquad
  [\fs_{\bar 0},\fs_{\bar 1}] \subset \fs_{\bar 1} \; ,
  \qquad\text{and}\qquad [\fs_{\bar 1},\fs_{\bar 1}] \subset \fs_{\bar
    0}~.
\end{equation}
Any such bracket on $\fs$ must obey the Jacobi identity in order to
define a Lie superalgebra. Each graded component of the Jacobi
identity is of type $[{\bar 0}{\bar 0}{\bar
  0}]$, $[{\bar 0}{\bar 0}{\bar 1}]$, $[{\bar 0}{\bar 1}{\bar 1}]$ or
$[{\bar 1}{\bar 1}{\bar 1}]$. The first three graded components can be
conceptualised as follows. The $[{\bar 0}{\bar 0}{\bar 0}]$ part says
that $\fs_{\bar 0}$ must be a Lie algebra with respect to $[\fs_{\bar
  0}, \fs_{\bar 0}]$, whence $\fs_{\bar 0} < \fC(M,g)$. The $[{\bar
  0}{\bar 0}{\bar 1}]$ part says that $[\fs_{\bar 0}, \fs_{\bar 1}]$
must define a representation of $\fs_{\bar 0}$ on $\fs_{\bar 1}$. The
$[{\bar 0}{\bar 1}{\bar 1}]$ part says that the symmetric bilinear map
defined by $[\fs_{\bar 1},\fs_{\bar 1}]$ must be equivariant with
respect to the $\fs_{\bar 0}$-action defined by $[\fs_{\bar 0},
\fs_{\bar 1}]$. Finally, the $[{\bar 1}{\bar 1}{\bar 1}]$ part, being
symmetric trilinear in its entries, is equivalent via polarisation to
the condition
\begin{equation}\label{eq:10dCSGodd3}
[ [ \epsilon,\epsilon], \epsilon ]=0~,
\end{equation}
for all $\epsilon \in \fs_{\bar 1}$. If $\fs$ is a Lie superalgebra,
notice that there exists a (possibly trivial) ideal $\fk_{\bar 0} :=
[\fs_{\bar 1},\fs_{\bar 1}] \lhd \fs_{\bar 0}$ and indeed $\fk :=
[\fs_{\bar 1},\fs_{\bar 1}] \oplus \fs_{\bar 1} \lhd \fs$ is a Lie
superalgebra ideal.

The Kosmann-Schwarzbach Lie derivative
\begin{equation}
  \label{eq:KSLD}
  {\hat \cL}_X = \nabla_X + \tfrac{1}{4} ( \nabla_\mu X_\nu )
  \Gamma^{\mu\nu} + \half \sigma_X \1~,
\end{equation}
along any $X \in \fC(M,g)$ (i.e., $\cL_X g = -2\sigma_X g$), defines a
natural conformally equivariant action of $\fC(M,g)$ on spinors. It is
therefore tempting to define 
\begin{equation}
  \label{eq:BFbracketKSLD}
  [ X , \epsilon ] = {\hat \cL}_X \epsilon~,
\end{equation}
for all $X \in \fs_{\bar 0}$ and $\epsilon \in \fs_{\bar 1}$. However, for any $\epsilon
\in \fs_{\bar 1}$, one finds that ${\hat \cL}_X \epsilon \in \fs_{\bar 1}$ (i.e., solving
\eqref{eq:10dCSGvacua}) only if
\begin{equation}\label{eq:LDHG}
  \Gamma_\mu ( 4 \Dirac \alpha_X + \beta_X ) \epsilon + 3 \beta_X
  \Gamma_\mu \epsilon =0~,
\end{equation}
where $\alpha_X = G_X + 3 \sigma_X$ and $\beta_X = \cL_X H + 2
\sigma_X H$, for all $X \in \fs_{\bar 0}$. Under a Weyl transformation
$(g,G,H) \mapsto ( \Omega^2 g , G + 3\,  \rd(\ln \Omega ) , \Omega^2
H)$ of the background, for any $X \in \fs_{\bar 0}$, it follows that
$\alpha_X \mapsto \alpha_X$ and $\beta_X \mapsto  \Omega^2 \beta_X$.
This implies that the condition \eqref{eq:LDHG} is Weyl-invariant. If
\eqref{eq:LDHG} is satisfied, the bracket \eqref{eq:BFbracketKSLD}
solves the $[{\bar 0}{\bar 0}{\bar 1}]$ Jacobi.

Now recall from above \eqref{eq:10dCSGvacuaSigmaxi} that any $\epsilon
\in \fs_{\bar 1}$ has Dirac current $\xi_\epsilon \in
\fC(M,g)$. Moreover, \eqref{eq:10dCSGvacuaSigmaxi} and
\eqref{eq:10dCSGvacuaLiexiH} are precisely the conditions
$\alpha_{\xi_\epsilon} =0$ and $\beta_{\xi_\epsilon} =0$ which, if
$\xi_\epsilon \in \fs_{\bar 0} < \fC(M,g)$, would ensure that
\eqref{eq:LDHG} is satisfied. With this in mind, let us now define the
$[\fs_{\bar 1},\fs_{\bar 1}]$ bracket such that
\begin{equation}
  \label{eq:FFbracket}
 [\epsilon,\epsilon] = \xi_\epsilon~,
\end{equation}
for all $\epsilon \in \fs_{\bar 1}$. Being symmetric bilinear in its
entries, the general $[\fs_{\bar 1},\fs_{\bar 1}]$ bracket follows via
the polarisation $\half ( \xi_{\epsilon + \epsilon^\prime} -
\xi_\epsilon - \xi_{\epsilon^\prime} ) = [ \epsilon , \epsilon^\prime
]$, for any $\epsilon , \epsilon^\prime \in \fs_{\bar 1}$. Given
\eqref{eq:LDHG}, it is straightforward to check that the symmetric
bilinear map defined by \eqref{eq:FFbracket} is indeed equivariant
with respect to the $\fs_{\bar 0}$-action defined by
\eqref{eq:BFbracketKSLD}, whence solving the $[{\bar 0}{\bar 1}{\bar
  1}]$ Jacobi. Furthermore, it follows using \eqref{eq:10dCSGvacuaCKV}
that
\begin{equation}\label{eq:10dCSodd3}
[ \xi_\epsilon , \epsilon ] = {\hat
  \cL}_{\xi_\epsilon} \epsilon =0~,
\end{equation}
for all $\epsilon \in \fs_{\bar 1}$, so the final $[{\bar 1}{\bar
  1}{\bar 1}]$ Jacobi is satisfied identically.

In summary, we have shown that the brackets defined by
\eqref{eq:BFbracketKSLD} and \eqref{eq:FFbracket} equip $\fs$ with the
structure of Lie superalgebra provided the condition \eqref{eq:LDHG}
is satisfied. Any such Lie superalgebra $\fs$ with $\fs_{\bar 0} <
\fC(M,g)$ maximal will be referred to as the {\emph{conformal symmetry
    superalgebra}} of $(M,g,G,H)$. By construction, a conformal
symmetry superalgebra $\fs$ must have $[\fs_{\bar 1},\fs_{\bar 1}]
\lhd \fs_{\bar 0} < \fC(M,g)$. The $\fs_{\bar 1}$-generated ideal $\fk
= [\fs_{\bar 1},\fs_{\bar 1}]  \oplus \fs_{\bar 1}$ of a conformal
symmetry superalgebra $\fs$ will be referred to as the
{\emph{conformal Killing superalgebra}} of $(M,g,G,H)$. It follows
that every bosonic supersymmetric background of conformal supergravity
in ten dimensions admits a conformal Killing superalgebra because
\eqref{eq:LDHG} is identically satisfied (as a consequence of
\eqref{eq:10dCSGvacuaSigmaxi} and \eqref{eq:10dCSGvacuaLiexiH}) for
all conformal Killing vectors in $[\fs_{\bar 1},\fs_{\bar 1}]$. Of
course, because the construction is manifestly Weyl-equivariant,
strictly speaking a conformal symmetry superalgebra is ascribed to a
conformal class of supersymmetric conformal supergravity backgrounds.

We will not attempt to obtain the general solution of \eqref{eq:LDHG}
though it will be useful to describe what happens for conformal
supergravity backgrounds which preserve more than half the maximal
amount of supersymmetry. A simple algebraic proof was given in
\cite[§3.3]{FigueroaO'Farrill:2012fp} that any bosonic supersymmetric
background of type I supergravity in ten dimensions which preserves
more than half the maximal amount of supersymmetry is necessarily
locally homogeneous. The same logic implies that any bosonic
supersymmetric background of conformal supergravity in ten dimensions
which preserves more than half the maximal amount of supersymmetry is
necessarily locally conformally homogeneous. In both cases, the trick
is to show that, for any given $x \in M$, the values at $x$ of all
(conformal) Killing vectors $\xi_\epsilon$ obtained by `squaring'
supersymmetry parameters $\epsilon$ span the tangent space $T_x M$
(i.e., the evaluation at $x$ of the squaring map
$\epsilon \mapsto \xi_\epsilon$ is surjective). Acting with
${\overline \epsilon} \Gamma^\mu$ on \eqref{eq:LDHG} implies
\begin{equation}
  \label{eq:LDHG2}
  \eL_{\xi_\epsilon} \alpha_X =0~, 
\end{equation}  
for all $\epsilon \in \fs_{\bar 1}$ and $X \in \fs_{\bar
  0}$.  Therefore, in this case, the condition \eqref{eq:LDHG2} says
that $\alpha_X$ must be (locally) constant, for all $X \in \fs_{\bar
  0}$.  The condition \eqref{eq:LDHG} then just says that, for any
vector field $Y$, the two-form $\iota_Y \beta_X$ must annihilate
$\epsilon$, for all $X \in \fs_{\bar 0}$ and $\epsilon \in \fs_{\bar
  1}$.  This means that the element $\iota_Y \beta_X \in \fspin(9,1)
\subset \Cl(9,1)$ annihilates a linear subspace of $\Delta^{(9,1)}_+$
of dimension $>8$ and hence by \cite[App.~B]{Gran:2007fu} (see also
\cite[Table~2]{AFS-groups}) it must vanish.  Thus, we have shown that
demanding \eqref{eq:LDHG} for all $\epsilon \in \fs_{\bar 1}$ with
$\dim \fs_{\bar 1} > 8$ implies
\begin{equation}
  \label{eq:LDHGHalfPlus}
  \rd \alpha_X =0 \qquad\text{and}\qquad \beta_X =0~,
\end{equation}  
for all $X \in \fs_{\bar 0}$, which then trivially implies~\eqref{eq:LDHG},
showing that they are equivalent.

Now consider the Weyl transformation defined above
\eqref{eq:10dCSGvacuaGzero}, which can be used to eliminate $G$. This
maps a supersymmetric conformal supergravity background $(M,g,G,H)$
with supersymmetry parameter $\epsilon$ to another supersymmetric
conformal supergravity background $(M, {\tilde g} = \Omega^2 g,
{\tilde G} = 0, {\tilde H} = \Omega^2 H)$ with supersymmetry parameter
${\tilde \epsilon} = \sqrt{\Omega} \epsilon$, where $\Omega =
\te^{-\Phi/3}$. If the conditions \eqref{eq:LDHGHalfPlus} are
satisfied then
\begin{equation}
  \label{eq:LDHGHalfPlusHat}
  \cL_X {\tilde g} = -\tfrac{2}{3} \alpha_X {\tilde g}
\qquad\text{and}\qquad \cL_X {\tilde H} =0~,
\end{equation}  
for all $X \in \fs_{\bar 0}$. The first condition in \eqref{eq:LDHGHalfPlusHat}
implies that every conformal Killing vector $X$ with respect to $g$ is
homothetic with respect to ${\tilde g}$ (since $\alpha_X$ is
constant), i.e., $\fC(M,g) = \fH(M,{\tilde g})$. Moreover, since
$\alpha_{\xi_\epsilon} =0$ for all $\epsilon \in \fs_{\bar 1}$, every conformal
Killing vector in $[\fs_{\bar 1},\fs_{\bar 1}]$ is a Killing vector with respect to
${\tilde g}$. In this case, $(M,g)$ being (locally) conformally
homogeneous implies that $(M,{\tilde g})$ is (locally) homogeneous.

\subsection{Maximally supersymmetric backgrounds}
\label{sec:10dCSGoffshellMaxSUSY} 

Maximally supersymmetric backgrounds are such that the connection
$\eD$ defined by equation \eqref{eq:10dCSGvacua} is flat.  Hence one
can determine the maximally supersymmetric backgrounds of conformal
supergravity in ten dimensions by solving the flatness equation which results by
abstracting $\epsilon$ from equation~\eqref{eq:10dCSGInt} and solving
the resulting equation for endomorphisms of the spinor bundle.

For maximally supersymmetric backgrounds of type I supergravity, the
condition $G \epsilon = \half H \epsilon$ implies $G=0$, $H=0$ and
\eqref{eq:10dCSGInt} then implies that the Riemann tensor must also
vanish. The only maximally supersymmetric background of type I
supergravity in ten dimensions is therefore locally isometric to
Minkowski space, which is Theorem~4 in
\cite{FigueroaO'Farrill:2002ft}.

For maximally supersymmetric backgrounds of conformal supergravity,
the flatness equation derived from \eqref{eq:10dCSGInt} implies
\begin{equation}
  \label{eq:MaxSUSYconditions}
  \nabla_\mu H_{\nu\rho\sigma} =0 \; , \qquad H_{\mu\nu[\rho}
  H_{\sigma\alpha\beta]} =0 \; , \qquad\text{and}\qquad H_{\mu\nu\rho}
  G_\sigma =0~.
\end{equation}
The last equation gives rise to two branches of solutions: those with
$H=0$ and those with $H\neq 0$ and hence $G=0$.  If $H=0$ then
\eqref{eq:MaxSUSYconditions} are trivially satisfied and the flatness
equation from \eqref{eq:10dCSGInt} is equivalent to
\begin{equation}
  \label{eq:MaxSUSYHzero}
  R_{\mu\nu\rho\sigma} = -\tfrac{2}{3} g_{\rho[\mu} ( \nabla_{\nu]}
  G_\sigma + \tfrac{1}{3} G_{\nu]} G_\sigma ) + \tfrac{2}{3}
  g_{\sigma[\mu} ( \nabla_{\nu]} G_\rho + \tfrac{1}{3} G_{\nu]} G_\rho
  ) + \tfrac{2}{9} g_{\rho[\mu} g_{\nu]\sigma} G_\alpha G^\alpha~.
\end{equation}
The condition \eqref{eq:MaxSUSYHzero} just says that the Riemann
tensor of the Weyl transformed metric $\te^{-2\Phi/3} g$ is zero.  In
other words, $g$ is conformally flat.

On the other hand, if $H\neq 0$, then the third condition in
\eqref{eq:MaxSUSYconditions} implies that $G=0$ and the flatness
equation derived from \eqref{eq:10dCSGInt} is equivalent to
\begin{equation}
  \label{eq:MaxSUSYGzero}
  R_{\mu\nu\rho\sigma} = \tfrac{1}{36} \left( 3 H_{\mu\nu}{}^\alpha
    H_{\rho\sigma\alpha} + g_{\rho[\mu} H_{\nu]}{}^{\alpha\beta}
    H_{\sigma\alpha\beta} - g_{\sigma[\mu} H_{\nu]}{}^{\alpha\beta}
    H_{\rho\alpha\beta} - \tfrac{1}{3} g_{\rho[\mu} g_{\nu]\sigma}
    H^{\alpha\beta\gamma} H_{\alpha\beta\gamma} \right)~,
\end{equation}
together with the first two conditions in
\eqref{eq:MaxSUSYconditions}.  The first of those conditions says that
$H$ is parallel with respect to the Levi-Civita connection $\nabla$
and, by equation \eqref{eq:MaxSUSYGzero}, so is the Riemann tensor of
$g$.  In other words, the background must be locally isometric to a
lorentzian symmetric space.  Now we shall classify maximally
supersymmetric backgrounds of conformal supergravity with $G=0$,
making use of several key techniques developed in
\cite{FigueroaO'Farrill:2002ft}.

The second condition in \eqref{eq:MaxSUSYconditions}, written in a
more invariant way, is
\begin{equation}
  \iota_X \iota_Y H \wedge H = 0~,
\end{equation}
for all vector fields $X,Y$.  This is none other than the family of
Plücker quadrics for $H$ (see, e.g., \cite[Ch.~1]{GrifHar}), which is
equivalent to $H$ being decomposable; that is, $H = \alpha \wedge
\beta \wedge \gamma$, for one-forms $\alpha, \beta, \gamma$.  Any
background of interest is therefore locally isometric to a
ten-dimensional lorentzian symmetric space $M$ equipped with a
parallel decomposable three-form $H$.  These conditions are quite
restrictive and solutions are distinguished according to whether 
the constant $\rn{H}^2 := \tfrac{1}{6} H_{\mu\nu\rho}
H^{\mu\nu\rho}$ is positive, negative or zero.  The geometric meaning
of this constant has to do with the metric nature of the tangent 3-planes
which $H$ defines: they can be either euclidean, lorentzian or
degenerate, according to whether $\rn{H}^2$ is positive, negative or
zero, respectively.   From \eqref{eq:MaxSUSYGzero}, it follows that
the constant scalar curvature of $g$ is $R = - \tfrac{1}{2} \rn{H}^2$.
The maximally supersymmetric backgrounds are summarised below (with
the scalar curvature of each $\AdS$ and $S$ factor denoted in
parenthesis).

\begin{itemize}
\item If $R>0$, $M = \AdS_3 (-\tfrac{4}{3} R) \times S^7
  (\tfrac{7}{3} R)$ with $H = \sqrt{2R} \, \vol_{\AdS_3}$.
\item If $R<0$, $M = \AdS_7 (\tfrac{7}{3} R) \times S^3
  (-\tfrac{4}{3} R)$ with $H = \sqrt{-2R} \, \vol_{S^3}$.
\item If $R=0$, $M = \CW_{10} (A)$ with $A =
  -\tfrac{\mu^2}{36} \diag (4,4,1,1,1,1,1,1)$ and $H = \mu
  \, \rd x^- \wedge \rd x^1 \wedge \rd x^2$. 
\end{itemize}

The background $\CW_{10} (A)$ denotes a ten-dimensional
Cahen--Wallach lorentzian symmetric space with metric
\begin{equation}
  \label{eq:CW10} 
  g = 2 \rd x^+ \rd x^- + \left( \sum_{a,b =1}^8 A_{ab} \, x^a x^b
  \right) ( \rd x^- )^2 + \sum_{a=1}^8 ( \rd x^a )^2~,
\end{equation}
in terms of local coordinates $(x^\pm,x^a)$.  For a general constant
symmetric matrix $A = ( A_{ab} )$, it follows that $g$ is conformally
flat only if $A$ is proportional to the identity matrix. Clearly this
is not the case for the particular $A$ which defines the maximally
supersymmetric background in the third item above (unless $\mu = 0$,
in which case $\CW_{10} (0) = \RR^{9,1}$).  Moreover, the
maximally supersymmetric backgrounds with $R\neq 0$ in the first two
items above are not conformally flat since, in each case, the constant 
sectional curvatures of the $\AdS$ and $S$ factors are not equal and
opposite (e.g., see (1.167) in \cite{Besse}).

It follows from \cite[§4]{Limits} that the Freund--Rubin backgrounds
$\AdS_3 \times S^7$ and $\AdS_7 \times S^3$ found above have two
distinct plane-wave (or Penrose--Güven) limits up to local isometry.
If the geodetic vector of the null geodesic along which we take the
limit is tangent to the anti-de~Sitter space, then the limit is flat,
whereas if the geodetic vector has a nonzero component tangent to the
sphere, the limit is isometric to the Cahen--Wallach background we
found above. Indeed, the ratio ($=4$, in this case) between the two
eigenvalues of the symmetric matrix $A$ defining the Cahen--Wallach
metric is the square of the ratio ($=2$, in this case) of the radii of
curvature of the 3- and 7-dimensional factors in the Freund--Rubin
geometry. This gives another proof that the Freund--Rubin backgrounds
are not conformally flat, since conformal flatness is a hereditary
property under the plane-wave limit \cite[§3.2]{Limits}, but the
$\CW_{10}(A)$ geometry above is not conformally flat for $\mu \neq 0$.


\subsubsection{Conformal symmetry superalgebras}
\label{sec:10dCSMaximalSUSYSuperalgebra}

Let us now investigate how the construction of conformal symmetry
superalgebras in Section~\ref{sec:10dCSKillingSuperalgebra} plays out
for the maximally supersymmetric conformal supergravity backgrounds we
have just classified.

For the maximally supersymmetric background with $H=0$,
$(M,{\tilde g})$ is locally isometric to $\RR^{9,1}$. The
supersymmetry condition \eqref{eq:10dCSGvacuaGzero} implies
$\fs_{\bar 1} \cong \Delta^{(9,1)}_+$ on $\RR^{9,1}$. Surjectivity of
the squaring map then implies
$[\fs_{\bar 1},\fs_{\bar 1}] \cong \RR^{9,1}$. For any
$\epsilon \in \Delta^{(9,1)}_+$,
${\hat \cL}_X \epsilon \in \Delta^{(9,1)}_+$ only if the conformal
Killing vector $X$ does not involve a special conformal transformation
in $\fC(\RR^{9,1}) \cong \fso(10,2)$. This is just as expected from
\eqref{eq:LDHGHalfPlus}, so that the associated conformal factor
$\sigma_X$ is constant. Thus, we must take
$\fs_{\bar 0} = \fH(\RR^{9,1}) < \fC(\RR^{9,1})$, which consists of
the obvious Poincaré transformations generated by
$\fK(\RR^{9,1}) \cong \fso(9,1) \ltimes \RR^{9,1}$ plus dilatation
generated by a proper homothetic conformal Killing vector
$\theta$. The Lie superalgebra obtained by restricting to
$\fK(\RR^{9,1}) \lhd \fH(\RR^{9,1})$ is isomorphic to the Poincaré
superalgebra in ten dimensions. The conformal symmetry superalgebra
$\fs$ merely appends $\theta$ to this Poincaré superalgebra, with the
additional bracket $[\theta,\epsilon] = \half \epsilon$, for all
$\epsilon \in \Delta^{(9,1)}_+$ (which implies
$[\theta , \xi_\epsilon ] = \xi_\epsilon$).

For all three maximally supersymmetric backgrounds with $H\neq 0$,
$G=0$ so $\fC(M,g) = \fH(M,g)$ because $\alpha_X = 3\sigma_X$ is
constant, for all $X \in \fC(M,g)$. Given any $X,Y \in \fH(M,g)$ with
$\sigma_X \neq 0$, then $Y - \frac{\sigma_Y}{\sigma_X} X \in
\fK(M,g)$. Hence, either $\fH(M,g) = \fK(M,g)$ or $\dim (
\fH(M,g) / \fK(M,g) ) =1$. For the two maximally supersymmetric
backgrounds with constant scalar curvature $R \neq 0$, a quick
calculation reveals that the scalar norm-squared of the Weyl tensor
$W$ of $g$ is ${\rn{W}}^2 = \frac{7}{36} R^2$.  Recall that the Weyl
tensor obeys $\cL_X W = -2 \sigma_X W$, for all $X \in \fC(M,g)$, so
$\eL_X {\rn{W}}^2 = 4 \sigma_X {\rn{W}}^2$. Hence, because in
this case ${\rn{W}}^2$ is a non-zero constant, it follows that
$\fH(M,g) = \fK(M,g)$.  The third maximally supersymmetric background
with $R=0$ does admit a proper homothetic conformal Killing vector so
$\dim ( \fH(M,g) / \fK(M,g) ) =1$.


\subsubsection{$\fs(\AdS_3 \times S^7 )$ and $\fs(\AdS_7 \times S^3 )$}
\label{sec:ConformalSuperalgebrasAdSS}

The preceding discussion has established that the only conformal
Killing vectors for these two geometries are Killing vectors.
Moreover, it is not difficult to prove that all such Killing vectors
correspond to Killing vectors on the individual $\AdS$ and $S$
factors.

The supersymmetry condition \eqref{eq:10dCSGvacuaGzero} reduces to a
pair of Killing spinor equations on the individual $\AdS$ and $S$
factors. In our conventions, a spinor $\psi$ on a
lorentzian/riemannian spin manifold $M$ is Killing if, for any vector
field $X$ on $M$, it obeys $\nabla_X \psi = \pm \tfrac{\kappa}{2} X
\psi$, for some real/imaginary constant $\kappa$. In the case at hand,
the Killing constants are given by $\kappa_{\AdS_3} = i \kappa_{S^3} =
\tfrac{1}{3} \sqrt{2 |R|}$ and $\kappa_{\AdS_7} = i \kappa_{S^7} =
\tfrac{1}{6} \sqrt{2 |R|}$.

Both $\AdS_m$ and $S^n$ can be described via the canonical quadric
embedding in (an open subset of) $\RR^{m-1,2}$ and $\RR^{n+1}$
respectively. Conversely, the flat metrics on both $\RR^{m-1,2}$ and
$\RR^{n+1}$ can be written as (lorentzian and riemannian) cone metrics
whose bases form the respective $\AdS_m$ and $S^n$ geometries. This
cone construction is particularly useful in describing Killing vectors
and Killing spinors on these geometries (see
\cite{FigueroaO'Farrill:1999va} for a review in a similar
context). Every Killing vector on the base lifts to a constant
two-form on the cone and vice versa. Thus
$\fK(\AdS_m) \cong \fso(m-1,2)$ ($\cong \wedge^2 \RR^{m-1,2}$ as a
vector space) and $\fK( S^n ) \cong \fso(n+1)$
($\cong \wedge^2 \RR^{n+1}$ as a vector space). Every Killing
spinor on the base lifts to a constant spinor on the cone and vice
versa. More precisely, if both $m$ and $n$ are odd, there is a
bijection between Killing spinors on the base and constant chiral
spinors on the cone. The Kosmann-Schwarzbach Lie derivative of a
Killing spinor along a Killing vector on the base lifts to the obvious
Clifford action of a constant two-form on a constant spinor on the
cone.

The cleanest way to discuss the explicit structure of $\fs(\AdS_3
\times S^7 )$ and $\fs(\AdS_7 \times S^3 )$ is as particular real
forms of the same complex Lie superalgebra $\fs^\CC$. The even part of
$\fs^\CC$ is $\fs_{\bar 0}^\CC = \fso_4 ( \CC ) \oplus \fso_8 ( \CC )$. The odd
part of $\fs^\CC$ is $\fs_{\bar 1}^\CC = \Delta_+^{(4 ,\CC )} \otimes
\Delta_+^{(8 ,\CC )}$, where $\Delta_+^{(4 , \CC )} \cong \CC^2$ and
$\Delta_+^{(8 , \CC )} \cong \CC^8$ denote the chiral spinor
representations of the respective $\fso_4 ( \CC )$ and $\fso_8 ( \CC
)$ factors in $\fs_{\bar 0}^\CC$. Let $\langle -,- \rangle$ denote the unique
(up to scale) $\fso_4 ( \CC )$-invariant skewsymmetric complex
bilinear form on $\Delta_+^{(4 , \CC )}$ and let $( -,- )$ denote the
unique (up to scale) $\fso_8 ( \CC )$-invariant symmetric complex
bilinear form on $\Delta_+^{(8 , \CC )}$.

Now fix a basis $( L_{AB} = - L_{BA} , M_{IJ} = - M_{JI} )$ for
$\fs_{\bar 0}^\CC$, where $A,B=1,2,3,4$ and $I,J = 1,\dots,8$. The brackets for
$\fs^\CC$ are as follows 
\begin{equation}\label{eq:SAdSSComplex}
  \begin{split}
    [ L_{AB} , L_{CD} ] &= - \delta_{AC} L_{BD} + \delta_{BC} L_{AD} +
    \delta_{AD} L_{BC} - \delta_{BD} L_{AC}~,\\
    [ M_{IJ} , M_{KL} ] &= - \delta_{IK} M_{JL} + \delta_{JK} M_{IL} +
    \delta_{IL} M_{JK} - \delta_{JL} M_{IK}~,\\
    [ L_{AB} , \psi \otimes \varphi ] &= \half \gamma_{AB} \psi
    \otimes \varphi\\ 
    [ M_{IJ} , \psi \otimes \varphi ] &= \half \psi \otimes
    \gamma_{IJ} \varphi~,\\
    [ \psi \otimes \varphi , \psi^\prime \otimes \varphi^\prime  ] &=
    \langle \psi , \gamma^{AB} \psi^\prime \rangle ( \varphi ,
    \varphi^\prime ) L_{AB} - \half \langle \psi , \psi^\prime \rangle
    ( \varphi , \gamma^{IJ} \varphi^\prime ) M_{IJ}~,
  \end{split}
\end{equation}
for all $\psi, \psi^\prime \in \Delta_+^{(4 , \CC )}$ and $\varphi,
\varphi^\prime \in\Delta_+^{(8 , \CC )}$, where $\{ \gamma_A \}$
generate $\CCl (4)$ and $\{ \gamma_I \}$ generate $\CCl (8)$. It is a
straightforward exercise to check that \eqref{eq:SAdSSComplex} obey
the graded Jacobi identities; although the $[{\bar 1}{\bar 1}{\bar 1}]$
component requires use of the following identities,
\begin{equation}
  \gamma^{AB} \psi \langle \gamma_{AB} \psi , - \rangle = -4 \psi
  \langle \psi , - \rangle
\end{equation}
and
\begin{equation}
  \gamma^{IJ} \varphi ( \gamma_{IJ} \varphi , - ) = -8 \varphi
  (\varphi , - ) + 8 ( \varphi , \varphi ) \1~,
\end{equation}
 which hold for all $\psi \in \Delta_+^{(4 , \CC )}$ and $\varphi
 \in\Delta_+^{(8 , \CC )}$.

As a vector space, $\fso_4 ( \CC ) \cong \wedge^2 \CC^4 \cong
\wedge_+^2 \CC^4 \oplus \wedge_-^2 \CC^4$, in terms of the
vector spaces $\wedge_\pm^2 \CC^4$ of (anti)self-dual two-forms on
$\CC^4$ which span each $\fsp_1 (\CC)$ factor in $\fso_4 ( \CC ) \cong
\fsp_1 (\CC) \oplus \fsp_1 (\CC)$. Let $\varepsilon \in \wedge^4
\CC^4$ with $\varepsilon_{1234} =1$ and let $\gamma = -\gamma_{1234}$
define the chirality matrix for $\CCl(4)$. It follows that
$\gamma_{AB} \gamma = \half \varepsilon_{ABCD} \gamma^{CD}$, so any
$\psi \in \Delta_+^{(4 , \CC )}$ defines a self-dual two-form $\langle
\psi , \gamma_{AB} \psi \rangle$. This implies that the bracket
defined by \eqref{eq:SAdSSComplex} of the $\fsp_1 (\CC) < \fso_4(\CC)$
spanned by $\wedge_-^2 \CC^4$ with every other element in $\fs^\CC$
is zero. The action of the other $\fsp_1 (\CC) < \fso_4(\CC)$ (spanned
by $\wedge_+^2 \CC^4$) on $\Delta_+^{(4 , \CC )}$ just corresponds
to the defining representation $\Delta^\CC$ of this $\fsp_1 (\CC)$.
Excluding the decoupled $\fsp_1 (\CC)$ factor from $\fs^\CC$ leaves a
simple complex Lie superalgebra that is isomorphic to $\fosp_{8|1}
(\CC)$ (a.k.a. $D(4,1)$ in the Kac classification \cite{Kac:1975}),
with even part $\fso_8 (\CC) \oplus \fsp_1 (\CC)$ and odd part
$\Delta_+^{(8 , \CC )} \otimes \Delta^{\CC}$. Thus, $\fs^\CC \cong
\fsp_1 (\CC) \oplus \fosp_{8|1} (\CC)$.

The real forms of all complex classical Lie superalgebras in
\cite{Kac:1975} were classified in \cite{Parker:1980}. Up to
isomorphism, the real forms of a given complex classical Lie
superalgebra are uniquely determined by the real forms of the complex
reductive Lie algebra which constitutes its even part. The even part
of $\fs^\CC$ is $\fs_{\bar 0}^\CC = \fso_4 ( \CC ) \oplus \fso_8 ( \CC )$ which
admits many non-isomorphic real forms. However, of these real forms,
only $\fso(2,2) \oplus \fso(8)$ and $\fso(6,2) \oplus \fso(4)$ are
isomorphic to the Lie algebra of isometries of $\AdS_3 \times S^7$ and
$\AdS_7 \times S^3$, respectively.  It is then straightforward to
deduce the associated real forms which describe their conformal
symmetry superalgebras $\fs$. The pertinent data is summarised in
Table~\ref{tab:OSPRealForms}.

\begin{table}[h!]
  \centering
  \setlength{\extrarowheight}{4pt}
  \caption{Data for admissible real forms of $\fs^\CC \cong \fsp_1
    (\CC) \oplus \fosp_{8|1} (\CC)$}
  \begin{tabular}{*4{>{$}c<{$}|}>{$}c<{$}}
    M & \fs_{\bar 0} & \fs_{\bar 1} & \text{type} & \fs \\[4pt]
    \hline
    \AdS_3 \times S^7 & \fso(2,2) \oplus \fso(8) & \Delta_+^{(2,2)} \otimes \Delta_+^{(8)} &  \RR & \fsl_2 (\RR) \oplus \fosp(8|2) \\[4pt] 
    \hline
    \AdS_7 \times S^3 & \fso(6,2) \oplus \fso(4) & [ \Delta_+^{(6,2)} \otimes \Delta_+^{(4)} ] & \HH & \fosp(6,2|1) \oplus \fsp (1) \\[4pt]
  \end{tabular}
  \label{tab:OSPRealForms}
\end{table}

We have opted for the more common physics notation to write the real
form $\fosp(8|2)$ rather than its perhaps more logical alias
$\fosp_{8|1} (\RR)$. The notation is that $\Delta_+^{(p,q)}$ denotes
the positive-chirality spinor representation of $\fso(p,q)$ when $p+q$
is even. As vector spaces, $\Delta_+^{(2,2)} \cong \RR^2$,
$\Delta_+^{(8)} \cong \RR^8$, $\Delta_+^{(6,2)} \cong \HH^4$ and
$\Delta_+^{(4)} \cong \HH$.  Given a pair of quaternionic
representations $W_1$ and $W_2$, which we think of as complex
representations equipped with invariant quaternionic structures $J_1$
and $J_2$, their tensor product $J_1 \otimes J_2$ defines a real
structure on $W_1 \otimes W_2$, where the tensor product is over
$\CC$.  This means that $W_1 \otimes W_2 \cong \CC \otimes_\RR [ W_1
\otimes W_2]$, where $[W_1 \otimes W_2]$ is a real representation
which can be identified with the subspace of real elements (i.e.,
fixed points of the real structure) in $W_1 \otimes W_2$.  Note that
$\fso(2,2) \cong \fsl_2 (\RR) \oplus \fsl_2 (\RR)$ with
$\Delta_+^{(2,2)} \cong \Delta \otimes \RR$, in terms of the defining
representation $\Delta$ of $\fsl_2 (\RR)$, while
$\fso(4) \cong \fsp(1) \oplus \fsp(1)$ with
$\Delta_+^{(4)} \cong \Delta^\prime \otimes \RR$, in terms of the
defining representation $\Delta^\prime$ of $\fsp(1)$, where we use
$\RR$ for the trivial real representation.


\subsubsection{$\fs(\CW_{10} (A))$}
\label{sec:ConformalSuperalgebrasPW}

To describe the conformal Killing vectors of the Cahen-Wallach
geometry \eqref{eq:CW10} with $A = -\tfrac{\mu^2}{36} \diag
(4,4,1,1,1,1,1,1)$, it is convenient to partition the indices
$a,b,\ldots$, which take values in $\{1,\dots,8\}$, into
$\alpha,\beta,\ldots  \in \{1,2\}$ and $i,j,\ldots \in
\{3,\dots,8\}$.

A basis of Killing vectors for this geometry is given by
\begin{equation}
  \label{eq:CWKV}
  \begin{aligned}[m]
    \xi &= \partial_+\\
    \zeta &= \partial_-\\
    J &= x^1 \partial_2 - x^2 \partial_1\\
    M_{ij} &= x^i \partial_j - x^j \partial_i\\
  \end{aligned}
  \qquad\qquad
  \begin{aligned}[m]
    q_\alpha &= \tfrac{3}{\mu} \sin ( \tfrac{\mu}{3}
    x^-) \partial_\alpha - x^\alpha \cos (\tfrac{\mu}{3}
    x^-) \partial_+\\
    q_i &= \tfrac{6}{\mu} \sin ( \tfrac{\mu}{6} x^- ) \partial_i - x^i
    \cos ( \tfrac{\mu}{6} x^- ) \partial_+\\
    p_\alpha &= \cos ( \tfrac{\mu}{3} x^- ) \partial_\alpha +
    \tfrac{\mu}{3} x^\alpha \sin ( \tfrac{\mu}{3} x^- ) \partial_+\\
    p_i &= \cos ( \tfrac{\mu}{6} x^- ) \partial_i + \tfrac{\mu}{6} x^i
    \sin ( \tfrac{\mu}{6} x^- ) \partial_+~.
  \end{aligned}
\end{equation}
Their non-vanishing Lie brackets are as follows    
\begin{equation}
  \label{eq:CWKVLB}
  \begin{aligned}[m]
    [ \zeta , q_\alpha ] &= p_\alpha\\
    [ \zeta , q_i ] &= p_i\\
    [ \zeta , p_\alpha ] &= - \tfrac{\mu^2}{9} q_\alpha\\
    [ \zeta , p_i ] &= - \tfrac{\mu^2}{36} q_i
  \end{aligned}
  \qquad\qquad
  \begin{aligned}[m]
    [ J , q_1 ] &= - q_2\\
    [ J , q_2 ] &= q_1\\
    [ J , p_1 ] &= - p_2\\
    [ J , p_2 ] &= p_1
  \end{aligned}
  \qquad\qquad
  \begin{aligned}[m]
    [ q_i , p_j ] &= \delta_{ij} \xi \\
    [ q_\alpha , p_\beta ] &= \delta_{\alpha\beta} \xi\\
    [ M_{ij} , q_k ] &= - \delta_{ik} q_j + \delta_{jk} q_i\\
    [ M_{ij} , p_k ] &= - \delta_{ik} p_j + \delta_{jk} p_i\\
  \end{aligned}
\end{equation}
in addition to
\begin{equation}
  \label{eq:CWKVLBtoo}
  [ M_{ij} , M_{kl} ] = - \delta_{ik} M_{jl} + \delta_{jk} M_{il} +
  \delta_{il} M_{jk} - \delta_{jl} M_{ik}~.
\end{equation}
The Killing vectors $(\xi , q , p)$ are generic for plane wave
geometries and we see from \eqref{eq:CWKVLB} that they form a
$17$-dimensional Lie subalgebra isomorphic to the Heisenberg algebra
$\fheis_8 (\RR)$. The Killing vectors $(J,M)$ span the Lie subalgebra
$\fso(2) \oplus \fso(6) < \fso(8)$ which stabilises $A$.

In total, notice that $\dim \fK(\CW_{10} (A)) = 34
= \dim \fK(\AdS_3 \times S^7) = \dim \fK(\AdS_7
\times S^3)$. However, $\CW_{10} (A)$ admits an additional
homothetic conformal Killing vector
\begin{equation}
  \label{eq:CWHomothety}
  \theta = 2 x^+ \partial_+ + x^a \partial_a~, 
\end{equation}
normalised such that $\sigma_\theta =-1$. Its non-vanishing Lie
brackets with $\fK(\CW_{10} (A))$ are
\begin{equation}
  \label{eq:CWHomothetyLB}
  [\xi , \theta ] = 2 \xi \; , \qquad [ q_a , \theta ] = q_a
  \qquad\text{and}\qquad [ p_a , \theta ] = p_a~.
\end{equation}
Thus we have obtained $\fs_{\bar 0}(\CW_{10} (A))$, defined with
respect to the basis $(\xi,q,p,\zeta,J,M,\theta) \in
\fH(\CW_{10} (A))$, subject to Lie brackets
\eqref{eq:CWKVLB} and \eqref{eq:CWHomothetyLB}.

The general solution of \eqref{eq:10dCSGvacuaGzero} on
$\CW_{10} (A)$ yields a supersymmetry parameter of the form
\begin{equation}\label{eq:CWSUSY}
\epsilon = \exp ( \tfrac{\mu}{4} x^- \bI ) \eta_+ + \half ( \Gamma_- +
\tfrac{\mu}{3} ( x^\alpha \Gamma_\alpha - \half x^i \Gamma_i ) \bI )
\, \exp ( \tfrac{\mu}{12} x^- \bI ) \eta_-~,
\end{equation}
in terms of $\bI = \Gamma_{12}$ and any pair of constant spinors
$\eta_\pm \in \Delta^{(9,1)}_\pm$ with $\Gamma_+ \eta_\pm =0$. It is perhaps
worth noting that a spinor of the form \eqref{eq:CWSUSY} on
$\CW_{10} (A)$ cannot be parallel with respect to the
Levi-Civita connection $\nabla$ unless it is identically zero
because any $\nabla$-parallel spinor $\psi$ on $\CW_{10}
(A)$ is necessarily constant with $\Gamma_+ \psi =0$.

Now let us adopt the shorthand notation $\epsilon =
\Psi(\eta_+,\eta_-)$ for any supersymmetry parameter of the form
\eqref{eq:CWSUSY}. Substituting \eqref{eq:CWKV},
\eqref{eq:CWHomothety} and \eqref{eq:CWSUSY} into the
Kosmann-Schwarzbach Lie derivative \eqref{eq:KSLD} yields the
following non-vanishing even-odd brackets for $\fs(\CW_{10}
(A))$
\begin{equation}
  \label{eq:CWEvenOddBrackets}
  \begin{aligned}[m]
    [ \zeta , \Psi(\eta_+,\eta_-) ] &= \tfrac{\mu}{4} \Psi( \bI
    \eta_+, \tfrac{1}{3} \bI \eta_-)\\
    [ J , \Psi(\eta_+,\eta_-) ] &= \half \Psi( \bI \eta_+,  \bI
    \eta_-)\\
    [ q_\alpha , \Psi(\eta_+,\eta_-) ] &= -\half \Psi( \Gamma_\alpha
    \eta_-,  0)\\
    [ p_\alpha , \Psi(\eta_+,\eta_-) ] &= \tfrac{\mu}{6} \Psi(
    \Gamma_\alpha \bI \eta_-,  0)
  \end{aligned}
  \qquad\qquad
  \begin{aligned}[m]
    [ \theta , \Psi(\eta_+,\eta_-) ] &= - \Psi( \eta_+, 0)\\
    [ M_{ij} , \Psi(\eta_+,\eta_-) ] &= \half \Psi( \Gamma_{ij} \eta_+, \Gamma_{ij} \eta_-)\\
    [ q_i , \Psi(\eta_+,\eta_-) ] &= -\half \Psi( \Gamma_i \eta_-,  0)\\
    [ p_i , \Psi(\eta_+,\eta_-) ] &= -\tfrac{\mu}{12} \Psi( \Gamma_i \bI\eta_-,  0)~.
  \end{aligned}
\end{equation}
Notice, in particular, that $\xi$ acts trivially on the Killing spinors.

Finally, substituting \eqref{eq:CWSUSY} into the squaring map
$\epsilon \mapsto \xi_\epsilon$ gives the odd-odd bracket
\begin{multline}
  \label{eq:CWOddOddBracket}
  [ \epsilon , \epsilon ] = (\overline{\eta}_+ \Gamma_- \eta_+) \xi
  - \half  (\overline{ \eta}_- \Gamma_- \eta_-) ( \zeta +
  \tfrac{\mu}{3} J) - \tfrac{\mu}{24} (\overline{ \eta}_- \Gamma_-
  \Gamma^{ij} \bI \eta_-)  M_{ij} \\
  -\tfrac{\mu}{3} (\overline{ \eta}_+ \Gamma_- \Gamma^{\alpha} \bI
  \eta_-)  q_\alpha +\tfrac{\mu}{6} (\overline{ \eta}_+ \Gamma_-
  \Gamma^{i} \bI  \eta_-)  q_i - (\overline{ \eta}_+ \Gamma_-
  \Gamma^{\alpha} \eta_-)  p_\alpha - (\overline{ \eta}_+ \Gamma_-
  \Gamma^{i} \eta_-)  p_i~.
\end{multline}

Thus we have obtained $\fs(\CW_{10} (A))$ and it is a simple matter to
confirm that the brackets defined by equations~\eqref{eq:CWKVLB}, \eqref{eq:CWKVLBtoo}, \eqref{eq:CWHomothetyLB}, \eqref{eq:CWEvenOddBrackets} and
\eqref{eq:CWOddOddBracket} indeed obey the Jacobi identities.  For any
$X \in \fH(\CW_{10} (A))$, $\alpha_X = 3\sigma_X$ is obviously
constant and one can check that $\beta_X =0$, as expected from
\eqref{eq:LDHGHalfPlus}, because the three-form
$H = \mu \, \rd x^- \wedge \rd x^1 \wedge \rd x^2$ obeys
$\cL_X H = -2\sigma_X H$.

\subsubsection{Contractions}
\label{sec:Contractions}

As we have seen, the Cahen--Wallach background is the plane-wave limit
of the Freund--Rubin backgrounds.  Therefore we might expect, based on what
happens in ten- and eleven-dimensional Poincaré supergravities
\cite{Limits,HatKamiSaka}, that $\fs(\CW_{10} (A))$ is a contraction
(in the sense of Inönü--Wigner) of $\fs(\AdS_3 \times S^7 )$ and
$\fs(\AdS_7 \times S^3 )$.  Indeed, it was precisely that observation
in \cite{NewIIB} which led to the identification of the maximally
supersymmetric plane-wave solutions of eleven-dimensional and IIB
supergravities as plane-wave limits of the corresponding Freund--Rubin
solutions in \cite{ShortLimits}.  We will give the details only for
the $\AdS_3 \times S^7$ Freund--Rubin background, and leave the
similar calculation for $\AdS_7 \times S^3$ to the imagination.

The contraction is easiest to describe in the following basis.  Let $\fs_{\bar 
  0} = \fso(2,2) \oplus \fso(8)$ be the even part of
$\fs(\AdS_3 \times S^7)$.  We will choose a basis $( P_\mu ,
L_{\mu\nu} )$ for $\fso(2,2)$ and $(P_m ,  M_{mn} )$ for
$\fso(8)$, where $\mu,\nu = 0,1,2$ and $m,n = 3,\dots,9$.  The Lie
brackets are given by
\begin{equation}
  \label{eq:so22so8}
  \begin{aligned}[m]
    [P_\mu, P_\nu] &= 4 L_{\mu\nu}\\
    [L_{\mu\nu},P_\rho] &= - \eta_{\mu\rho} P_\nu + \eta_{\nu\rho} P_\mu 
  \end{aligned}
  \qquad\qquad
  \begin{aligned}[m]
    [P_m, P_n] &= -M_{mn}\\
    [M_{mn},P_p] &= - \delta_{mp} P_n + \delta_{np} P_m
  \end{aligned}
\end{equation}
and
\begin{equation}
  \label{eq:so22so8too}
  \begin{split}
    [L_{\mu\nu},L_{\rho\sigma}] &= - \eta_{\mu\rho} L_{\nu\sigma} +\eta_{\nu\rho}
    L_{\mu\sigma}  + \eta_{\mu\sigma}
    L_{\nu\rho} -
    \eta_{\nu\sigma} L_{\mu\rho} \\
    [M_{mn},M_{pq}] &= - \delta_{mp} M_{nq} + \delta_{np} M_{mq} + \delta_{mq} M_{np} -
    \delta_{nq} M_{mp}~,
  \end{split}
\end{equation}
where $\eta = \diag(-1,+1,+1)$.  Let $\Psi : \Delta_+^{(9,1)} \to
\fs_{\bar 1}$ be a vector space isomorphism.  The even-odd brackets
of the conformal symmetry superalgebra of $\AdS_3 \times S^7$ are given
by
\begin{equation}
  \label{eq:MQ}
  [L_{\mu\nu}, \Psi(\varepsilon)] = \Psi(\tfrac12 \Gamma_{\mu\nu} \varepsilon)
  \qquad\text{and}\qquad [M_{mn}, \Psi(\varepsilon)] = \Psi(\tfrac12
  \Gamma_{mn} \varepsilon)~,
\end{equation}
\begin{equation}
  \label{eq:PQ}
  [P_\mu, \Psi(\varepsilon)] = \Psi(\Gamma_\mu \nu \varepsilon)
  \qquad\text{and}\qquad [P_m, \Psi(\varepsilon)] = - \tfrac12
  \Psi(\Gamma_m \nu \varepsilon)~,
\end{equation}
in terms of the $\Cl(9,1)$ gamma matrices and where $\varepsilon \in
\Delta_+^{(9,1)}$ and $\nu = \Gamma_{012}$.  Finally, the odd-odd
brackets are given by
\begin{equation}
  \label{eq:QQ}
  [\Psi(\varepsilon),\Psi(\varepsilon)] = (\overline{\varepsilon}\Gamma^\mu
  \varepsilon) P_\mu + (\overline{\varepsilon}\Gamma^m \varepsilon) P_m   -
  (\overline{\varepsilon}\Gamma^{\mu\nu}\nu \varepsilon) L_{\mu\nu} +
  \tfrac12   (\overline{\varepsilon}\Gamma^{mn} \nu \varepsilon) M_{mn}~.
\end{equation}
We now decompose $\varepsilon = \varepsilon_+ + \varepsilon_-$, with
$\varepsilon_\pm \in \ker \Gamma_\pm$, and expand the above Lie
bracket as follows, where the indices $\alpha,\beta \in \{1,2\}$ and
$i,j \in \{3,\dots,8\}$:
\begin{equation}
  \begin{split}
    [\Psi(\varepsilon),\Psi(\varepsilon)] &= (\overline{\varepsilon}_+\Gamma_- \varepsilon_+) (P_+ - L_{12}) - \tfrac14
    (\overline{\varepsilon}_+\Gamma^{ij} \bI \Gamma_- \varepsilon_+) M_{ij}\\
    & {} + (\overline{\varepsilon}_- \Gamma_+ \varepsilon_-) (P_- + 2 L_{12})
    + \tfrac12 (\overline{\varepsilon}_- \Gamma^{ij} \bI \Gamma_+ \varepsilon_-) M_{ij}\\
    & {} + 2 (\overline{\varepsilon}_+\Gamma^\alpha \varepsilon_-) P_\alpha + 2 (\overline{\varepsilon}_+\Gamma^i
    \varepsilon_-) P_i + 4 (\overline{\varepsilon}_+\Gamma^\alpha \bI
    \varepsilon_-) L_{0\alpha} - 2 (\overline{\varepsilon}_+ \Gamma^i \bI
    \varepsilon_-) M_{9i}~,
  \end{split}
\end{equation}
where we have defined $P_+ = \tfrac12 (P_9 + P_0)$ and $P_- = P_9 -
P_0$.

Let us now define a real $\ZZ_2$-graded vector space
$E = E_{\bar 0} \oplus E_{\bar 1}$, where $E_{\bar 0}$ is spanned by
the symbols $(\xi',\zeta', J', M'_{ij}, p'_\alpha,q'_\alpha, p'_i,
q'_i)$ for $\alpha,\beta = 1,2$ and $i,j=3,\dots,8$ and $E_{\bar 1}$
is the isomorphic image of $\Psi': \Delta_+^{(9,1)} \to E_{\bar 1}$.
We will define a family $\Upsilon_t : E \to \fs$ of even $\ZZ_2$-graded
linear maps 
by extending the following maps linearly:
\begin{equation}
  \label{eq:contraction}
  \begin{aligned}[m]
    \Upsilon_t(\xi') &= \tfrac{\mu}{12} t^2 \left(P_9 + P_0\right)\\
    \Upsilon_t(\zeta') &= \tfrac{\mu}{6} \left(P_9 - P_0\right)\\
    \Upsilon_t(J') &= L_{12}\\
    \Upsilon_t(M'_{ij}) &= M_{ij}
  \end{aligned}
  \qquad\qquad
  \begin{aligned}[m]
    \Upsilon_t(p'_\alpha) &= \tfrac{\mu}{6} t P_\alpha\\
    \Upsilon_t(p'_i) &= \tfrac{\mu}{6} t P_i\\
    \Upsilon_t(q'_\alpha) &= t L_{0\alpha}\\
    \Upsilon_t(q'_i) &= t M_{9i}
  \end{aligned}
\end{equation}
and where, for $\varepsilon \in \Delta_+^{(9,1)}$,
\begin{equation}
  \label{eq:contractiontoo}
  \Upsilon_t(\Psi'(\varepsilon)) =
  \begin{cases}
    \lambda t \Psi(\varepsilon)~, & \text{if $\varepsilon \in
      \ker\Gamma_+$}\\
    \lambda \Psi(\varepsilon)~, & \text{if $\varepsilon \in
      \ker\Gamma_-$}~,
  \end{cases}
\end{equation}
with $\lambda^2 = \frac{\mu}{6}$.  (We tacitly assume $\mu > 0$, but
in fact the factor $\mu$ is inessential and can always be taken to be
$1$, if nonzero.)  It is clear by inspection that $\Upsilon_t$
defines a vector space isomorphism for any $t \neq 0$.  For
definiteness, let us take $t > 0$.  We may define a family of Lie
brackets $[-,-]_t$ on $E$ by transporting the Lie bracket on
$\fs$ via $\Upsilon_t$:
\begin{equation}
  [x,y]_t := \Upsilon_t^{-1} [\Upsilon_t(x), \Upsilon_t(y)]~,
\end{equation}
for $x,y \in E$. By construction, for every $t > 0$, $(E,[-,-]_t)$ and
$(\fs,[-,-])$ are isomorphic Lie superalgebras.  If the limit
$t \to 0$ exists, then $(E,[-,-]_0)$ defines a Lie superalgebra, which
is then a \emph{contraction} of $(\fs,[-,-])$.  One checks that for
the map $\Upsilon_t$ defined in equations~\eqref{eq:contraction} and
\eqref{eq:contractiontoo}, the limit $t \to 0$ of the $[-,-]_t$
bracket does exist and that the resulting bracket is precisely the one
defined by equations~\eqref{eq:CWKVLB}, \eqref{eq:CWKVLBtoo}, \eqref{eq:CWEvenOddBrackets} (without the $\theta$ bracket) and
\eqref{eq:CWOddOddBracket}, once we remove the primes from the
symbols, and identify $\eta_+ = \varepsilon_+$ and
$\eta_- = \Gamma_+ \varepsilon_-$.

Let us illustrate this with some examples.  Firstly, let us consider the
bracket $[q'_i, p'_j]$, which is given by
\begin{equation}
  \begin{split}
    [q'_i, p'_j] &= \lim_{t \to 0} \Upsilon_t^{-1}
    [\Upsilon_t(q'_i), \Upsilon_t(p'_j)]\\
    &= \lim_{t \to 0} \Upsilon_t^{-1}
    [t M_{9i}, \tfrac\mu6 t P_j]\\
    &= \lim_{t \to 0} \tfrac\mu6 t^2
    \Upsilon_t^{-1}\delta_{ij} P_9\\
    &= \lim_{t \to 0} \delta_{ij} \tfrac\mu6 \left(\tfrac6\mu \xi' +
      \tfrac{3t^2}{\mu} \zeta'\right)\\
    &= \delta_{ij} \xi'~,
  \end{split}
\end{equation}
which agrees with equation \eqref{eq:CWKVLB}.  Next we consider the
bracket $[p'_\alpha, \Psi'(\varepsilon_-)]$, for $\varepsilon_- \in
\ker \Gamma_-$, given by
\begin{equation}
  \begin{split}
    [p'_\alpha, \Psi'(\varepsilon_-)] &= \lim_{t \to 0} \Upsilon_t^{-1}
    [\Upsilon_t(p'_\alpha), \Upsilon_t(\Psi'(\varepsilon_-))]\\
    &= \lim_{t \to 0} \Upsilon_t^{-1}
    [\tfrac\mu6 t P_\alpha,  \lambda \Psi(\varepsilon_-)]\\
    &= \lim_{t \to 0} \tfrac{\mu\lambda}6 \Upsilon_t^{-1} t 
    \Psi(\Gamma_\alpha \nu \varepsilon_-)\\
    &= \tfrac\mu6 \Psi'(\Gamma_\alpha I \Gamma_+ \varepsilon_-)~,
  \end{split}
\end{equation}
which shows that $\eta_- = \Gamma_+ \varepsilon_-$ for agreement with
equation~\eqref{eq:CWEvenOddBrackets}.  Next, we consider the bracket
$[\Psi'(\varepsilon_+),\Psi'(\varepsilon_+)]$, where
$\varepsilon_+ \in \ker\Gamma_+$, whose contraction is
\begin{equation}
  \begin{split}
    [\Psi'(\varepsilon_+),\Psi'(\varepsilon_+)] &= \lim_{t \to 0} \Upsilon_t^{-1}
    [\Upsilon_t(\Psi'(\varepsilon_+)), \Upsilon_t(\Psi'(\varepsilon_+))]\\
    &= \lambda^2 \lim_{t \to 0} t^2 \Upsilon_t^{-1}
    [\Psi(\varepsilon_+), \Psi(\varepsilon_+)]\\
    &= \tfrac\mu6 \lim_{t \to 0} t^2 \Upsilon_t^{-1} \left((\overline{\varepsilon}_+\Gamma_- \varepsilon_+)  (P_+ - L_{12}) - \tfrac12
    (\overline{\varepsilon}_+\Gamma^{ij} \bI \Gamma_- \varepsilon_+) M_{ij}\right)\\
    &=  \tfrac\mu6  \lim_{t \to 0} \left((\overline{\varepsilon}_+\Gamma_- \varepsilon_+) (\tfrac6\mu \xi' -
      t^2 J') - \tfrac12 (\overline{\varepsilon}_+\Gamma^{ij} \bI \Gamma_- \varepsilon_+) t^2
      M'_{ij}\right)\\
    &= (\overline{\varepsilon}_+\Gamma_- \varepsilon_+) \xi'~,
  \end{split}
\end{equation}
and $[\Psi'(\varepsilon_-),\Psi'(\varepsilon_-)]$, given by
\begin{equation}
  \begin{split}
    [\Psi'(\varepsilon_-),\Psi'(\varepsilon_-)] &= \lim_{t \to 0} \Upsilon_t^{-1}
    [\Upsilon_t(\Psi'(\varepsilon_-)), \Upsilon_t(\Psi'(\varepsilon_-))]\\
    &= \lambda^2 \lim_{t \to 0} \Upsilon_t^{-1}
    [\Psi(\varepsilon_-), \Psi(\varepsilon_-)]\\
    &=  \tfrac\mu6 \lim_{t \to 0} \Upsilon_t^{-1}
    \left((\overline{\varepsilon}_- \Gamma_+ \varepsilon_-) (P_- + 2 L_{12})
      + \tfrac12 (\overline{\varepsilon}_- \Gamma^{ij} \bI \Gamma_+ \varepsilon_-)
      M_{ij}\right)\\
    &=  \tfrac\mu6  \lim_{t \to 0} \left((\overline{\varepsilon}_- \Gamma_+ \varepsilon_-) (\tfrac6\mu
      \zeta' + 2 J') + \tfrac12 (\overline{\varepsilon}_- \Gamma^{ij} \bI \Gamma_+ \varepsilon_-)
      M'_{ij}\right)\\
    &= (\overline{\varepsilon}_- \Gamma_+
      \varepsilon_-) (\zeta' + \tfrac\mu3 J') + \tfrac\mu{12}
      (\overline{\varepsilon}_- \Gamma^{ij} \bI \Gamma_+
      \varepsilon_-) M'_{ij}~,
  \end{split}
\end{equation}
which agree with the first line of equation
\eqref{eq:CWOddOddBracket}, again using $\eta_- = \Gamma_+
\varepsilon_-$.

Finally, we should remark that the infinitesimal homothety $\theta$ of
the Cahen--Wallach background is not inherited from the Freund--Rubin
backgrounds via the plane-wave limit, hence we are not obtaining the
full conformal symmetry superalgebra as a contraction.  Of course,
this is not unexpected.



\subsubsection{Maximal superalgebras}
\label{sec:MaximalSA}

In \cite{FHJMS-MESA} the notion of the \emph{maximal superalgebra} of
a supergravity background was introduced, generalising to non-flat
backgrounds the M-algebra of \cite{TownsendMTfS}.  Given a
supergravity background with Killing superalgebra $\fk = \fk_{\bar 0}
\oplus \fk_{\bar 1}$, the maximal superalgebra (should it exist) is
defined to be Lie superalgebra $\fm = \fm_{\bar 0} \oplus \fm_{\bar
  1}$, satisfying the following properties
\begin{enumerate}
\item $\fm_{\bar 1} = \fk_{\bar 1}$ and $\fk_{\bar 0}$ is a Lie
  subalgebra of $\fm_{\bar 0}$;
\item the odd-odd bracket is an isomorphism $\odot^2\fm_{\bar 1} \cong
  \fm_{\bar 0}$; and
\item the projection $\odot^2 \fm_{\bar 1} \to \fk_{\bar 0}$ coincides
  with the odd-odd bracket of $\fk$ and the restriction to $\fk_{\bar
    0}$ of the bracket $\fm_{\bar 0} \otimes \fm_{\bar 1} \to
  \fm_{\bar 1}$ is the $\fk$-bracket.
\end{enumerate}
In other words, writing $\fm_{\bar 0} = \fk_{\bar 0} \oplus
\fz_{\bar 0}$, then the $\fm$-brackets $[\fk_{\bar 0},\fk_{\bar 0}]$,
$[\fk_{\bar 0},\fm_{\bar 1}]$ and the $\fk_{\bar 0}$-component of
$[\fm_{\bar 1}, \fm_{\bar 1}]$ are, respectively, the $[\fk_{\bar 0},\fk_{\bar 0}]$,
$[\fk_{\bar 0},\fk_{\bar 1}]$ and $[\fk_{\bar 1}, \fk_{\bar 1}]$
brackets of $\fk$, and only the brackets involving $\fz_{\bar 0}$ are
genuinely new.

As reviewed in \cite{FHJMS-MESA}, it follows from
\cite[App.~A]{KamimuraSakaguchi} that any Lie superalgebra satisfying
(2) --- in particular, the maximal superalgebra of a background --- is
uniquely determined by some
$\omega \in \left(\wedge^2\fm_{\bar 1}^*\right)^{\fm_0} \subset
\left(\wedge^2\fk_{\bar 1}^*\right)^{\fk_0}$.
Indeed, by (2) above, if $Q_a$ is a basis for $\fm_{\bar 1}$,
$Z_{ab} := [Q_a, Q_b]$ is a basis for $\fm_{\bar 0}$ and the Lie
brackets in this basis are given by
\begin{equation}
  \label{eq:maxalg}
  \begin{aligned}[m]
    [Z_{ab}, Q_c] &= \omega_{ac} Q_b + \omega_{bc} Q_a \\
    [Z_{ab}, Z_{cd}] &= \omega_{ac} Z_{bd} + \omega_{bc} Z_{ad} + \omega_{ad} Z_{bc} + \omega_{bd} Z_{ac}~,
  \end{aligned}
\end{equation}
where $\omega_{ab} := \omega(Q_a,Q_b)$ and where the second equation
follows from the first, using the Jacobi identity and the definition
of $Z_{ab}$.  In this section we explore the maximal superalgebras of
the maximally supersymmetric backgrounds.

First of all, we show that the Cahen--Wallach background does not
admit a maximal superalgebra. The proof is virtually identical to the
one for the Cahen--Wallach vacua of eleven-dimensional and type IIB
Poincaré supergravities in \cite{FHJMS-MESA}. As explained in
\cite[§3]{FHJMS-MESA}, for any maximal superalgebra $\fm$,
$\fk_{\bar 0}$ acts trivially on the radical $\fk_{\bar 1}^\perp$ of
the skew-symmetric bilinear form $\omega$ characterising $\fm$. Now,
inspecting equation~\eqref{eq:CWEvenOddBrackets} we see that
$\zeta = \partial_-$ acts semisimply on $\fk_{\bar 1}$ with nonzero
eigenvalues, so that $\fk_{\bar 1}^{\fk_{\bar 0}} = 0$. Therefore
$\omega$, having trivial radical, must be symplectic and hence, from
equation~\eqref{eq:maxalg}, it follows that $\fm$ must have trivial
centre. But now notice that $\xi = \partial_+$ acts trivially on
$\fm_{\bar 1}$ and hence it is central in $\fm$, thus contradicting
the existence of a maximal superalgebra for the Cahen--Wallach
background.

Next we discuss the two maximally supersymmetric Freund--Rubin
backgrounds $\AdS_7 \times S^3$ and $\AdS_3 \times S^7$.  Here it is
convenient to again think of their Killing superalgebras as different
real forms of the same complex Lie superalgebra.  At the same time we
must make a distinction between the symmetry superalgebra $\fs^\CC$
and the Killing superalgebra, which is the ideal $\fk^\CC$ of
$\fs^\CC$ generated by $\fk_{\bar 1}^\CC$.  For the Freund--Rubin
backgrounds, $\fs^\CC$ is strictly larger, containing a simple ideal
isomorphic to $\fsl_2 (\CC)$, which acts trivially on the Killing
spinors.

In this case, we have
$\fk^\CC_{\bar 0} \cong \fsl_2(\CC) \oplus \fso_8(\CC)$ and
$\fk^\CC_{\bar 1} \cong \Delta^\CC \otimes \Delta_+^{(8,\CC)}$ as an
$\fk^{\CC}_{\bar 0}$-module, where $\Delta^\CC$ is the defining
representation of $\fsl_2(\CC)$ and the tensor product is over $\CC$.
There is precisely one invariant skew-symmetric bilinear form (up to
scale) on $\fk_{\bar 1}^\CC$: it is the product of the
$\fsl_2(\CC)$-invariant complex symplectic structure
$\langle-,-\rangle$ on $\Delta^\CC$ and the $\fso_8(\CC)$-invariant
complex orthogonal structure $(-,-)$ on $\Delta_+^{(8,\CC)}$. It is
clearly nondegenerate, hence complex symplectic.  Therefore if
$\fk^\CC$ admits a maximal superalgebra, it has to be the
complexification $\fosp(1|16)^\CC$ of $\fosp(1|16)$.   On the other
hand, it is shown in \cite{Parker:1980} that $\fosp(1|16)$ is the
unique real form of $\fosp(1|16)^\CC$, so if `maximisation' were to
commute with complexification, we would conclude that the maximal
superalgebras of the Freund--Rubin backgrounds would be isomorphic to
$\fosp(1|16)$.  We do not have such a result at our disposal and it is
unlikely that such a general result actually exists since not every
Lie superalgebra can be `maximised', as illustrated by the Killing
superalgebra of the Cahen--Wallach backgrounds.  This means we need to
work harder.

We start by showing that the maximal superalgebra of $\fk^\CC$ is
indeed isomorphic to $\fosp(1|16)^\CC$, following the construction in
\cite{FHJMS-MESA}, mutatis mutandis.  First of all, let us decompose
the symmetric square of $\fm_{\bar 1}^\CC$ into irreducible
representations of $\fk_{\bar 0}^\CC$:
\begin{equation}
  \begin{split}
    \odot^2\fm_{\bar 1}^\CC &= \odot^2 \left(\Delta^\CC \otimes
      \Delta_+^{(8,\CC)}\right)\\
    &= \left( \odot^2 \Delta^\CC \otimes \odot^2 \Delta_+^{(8,\CC)}\right)
    \oplus \left(\wedge^2 \Delta^\CC \otimes \wedge^2 \Delta_+^{(8,\CC)}
    \right)\\
    &= \left( \odot^2 \Delta^\CC \otimes \left(\CC \oplus \odot_0^2
      \Delta_+^{(8,\CC)}\right)\right) \oplus \left(\CC \otimes \wedge^2
      \Delta_+^{(8,\CC)}\right)\\
    &= \odot^2 \Delta^\CC  \oplus \wedge^2  \Delta_+^{(8,\CC)} \oplus
    \left(\odot^2\Delta^\CC \otimes \odot_0^2  \Delta_+^{(8,\CC)}\right)\\
    &\cong \fsl_2(\CC) \oplus \fso_8(\CC) \oplus \fz_{\bar 0}^\CC~,
  \end{split}
\end{equation}
which defines $\fz_{\bar 0}^\CC$.  Except for the $\fz_{\bar 0}^\CC$, this
is precisely the odd-odd bracket of $\fk^\CC$.  It follows from the
discussion in \cite[§4.3]{FHJMS-MESA} that the action of $\fk_{\bar
  0}$ on $\fk_{\bar 1}$, when viewed through the lens of the cone
construction, is via the Clifford action of the parallel 2-forms on
the cones of $\AdS_p$ and $S^q$ to which the special Killing 1-forms
in $\fk_{\bar   0}$ lift. Therefore we extend this Clifford action to
all of $\fm_{\bar 0}$, which also lift as parallel forms to the cones.
This complexifies and gives the following construction of $\fm^\CC$;
although we prefer to use a different basis, which unfortunately
obscures the embedding of $\fk_{\bar 0}^\CC$ into $\fm_{\bar 0}^\CC$.

If $\psi_1\otimes \varphi_1,\psi_2 \otimes \varphi_2 \in \fm_{\bar
  1}^\CC$, their symplectic inner product is given by
\begin{equation}
  \omega(\psi_1\otimes \varphi_1,\psi_2 \otimes \varphi_2) =
  \left<\psi_1,\psi_2\right> \left(\varphi_1,\varphi_2\right)~.
\end{equation}
We define the following rank-1 endomorphisms of $\Delta^\CC$ and
$\Delta_+^{(8,\CC)}$:
\begin{equation}
  \psi_1\overline\psi_2 := \left<\psi_2,-\right>\psi_1
  \qquad\text{and}\qquad
 \varphi_1\overline\varphi_2 := \left(\varphi_2,-\right)\varphi_1~,
\end{equation}
and we define the odd-odd bracket in $\fm^\CC$ via
\begin{equation}
  [\psi_1\otimes \varphi_1,\psi_2 \otimes \varphi_2] =
  \psi_1\overline\psi_2 \otimes  \varphi_1\overline\varphi_2  + 
  \psi_2\overline\psi_1 \otimes  \varphi_2\overline\varphi_1~.
\end{equation}
By a judicious use of the Fierz identities, the rank-1 endomorphisms
above can be expressed in terms of the standard basis for the Clifford
algebra in terms of exterior forms, and in this way clarify the
embedding $\fk_{\bar 0}^\CC \subset \fm_{\bar 0}^\CC$, but we have no
need to do that.  The action of $\fm_{\bar 0}^\CC$ on $\fm_{\bar
  1}^\CC$ is given simply by the Clifford action, which is
\begin{equation}
  \begin{split}
    [\psi_1\overline\psi_2 \otimes  \varphi_1\overline\varphi_2  + 
  \psi_2\overline\psi_1 \otimes  \varphi_2\overline\varphi_1, \psi_3
  \otimes \varphi_3 ] &= \left<\psi_2,\psi_3\right>
  \left(\varphi_2,\varphi_3\right)\psi_1 \otimes \varphi_1 + 
  \left<\psi_1,\psi_3\right> \left(\varphi_1,\varphi_3\right)\psi_2
  \otimes \varphi_2\\
  &= \omega(\psi_2 \otimes \varphi_2,\psi_3 \otimes \varphi_3) \psi_1
  \otimes \varphi_1\\
  & \quad {} + \omega(\psi_1 \otimes \varphi_1,\psi_3 \otimes
  \varphi_3) \psi_2 \otimes \varphi_2~,
  \end{split}
\end{equation}
which agrees with the first equation in \eqref{eq:maxalg},
showing that indeed $\fm^\CC \cong \fosp(1|16)^\CC$.

How about the maximal subalgebras of $\fk(\AdS_3 \times S^7)$ and
$\fk(S^3 \times \AdS_7)$?  Let $\fk$ be one of these Killing
superalgebras.  It is a real form of $\fk^\CC$, so in particular
$\fk_{\bar 1}$ is the real subspace of  $\fk_{\bar 1}^\CC$ defined by
a $\fk_{\bar 0}$-invariant conjugation.  Now consider $\fk_{\bar 1}$
as a real subspace of $\fm_{\bar 1}^\CC$.  It generates a real
subalgebra of $\fm^\CC$, which satisfies property (2) of a maximal
subalgebra because the restriction to $\fk_{\bar 1}$ of the odd-odd
bracket is an isomorphism onto its image.  This means that the
brackets are of the form \eqref{eq:maxalg} with $\omega$ being the
restriction of the complex-symplectic form on $\fm_{\bar 1}^\CC$ to
the real subspace $\fk_{\bar 1}$.  The other properties for a maximal
subalgebra are satisfied because $\fk^\CC$ \emph{is} the
complexification of $\fk$.  Therefore we see that $\fk$ admits a
maximal superalgebra, but to identify it we need to understand the
restriction of $\omega$ to $\fk_{\bar 1}$.  It pays to be a little bit
more general.

Let $(E,\omega)$ be a complex symplectic vector space.  Let $\fg$ be a
Lie algebra, whose complexification $\fg^\CC$ acts on $E$ preserving
$\omega$.  Now suppose that $c$ is a $\fg$-invariant conjugation on
$E$ and let $E^\RR$ be its fixed (real) subspace; that is,
$E = E^\RR \otimes_\RR \CC$.  Because $c$ is $\fg$-invariant,
$\fg$ acts on $E^\RR$.  Now, $\omega$ restricts to a real
skewsymmetric bilinear form $\omega^\RR$ on $E^\RR$.  Since
$\omega$ is $\fg^\CC$-invariant, it is in particular also
$\fg$-invariant and hence so is $\omega^\RR$.  Its radical, therefore,
is a $\fg$-submodule of $E^\RR$.  Now suppose that $E^\RR$ is
irreducible as a $\fg$-module.  Then the radical of $\omega^\RR$ must
either be trivial, in which case $\omega^\RR$ is a symplectic form, or
it must be all of $E^\RR$, in which case $\omega^\RR = 0$.

Now let us apply this to our situation, with the rôle of $(E,\omega)$
played by $(\fk_{\bar 1}^\CC, \omega)$.  We have that $\fk_{\bar 1}$
is an irreducible module of $\fk_{\bar 0}$, so that the restriction of
$\omega$ to $\fk_{\bar 1}$ is either symplectic or zero.  But it
cannot be zero, because otherwise $\fk_{\bar 0}$ would be central and in
particular, an abelian Lie algebra.  Therefore we conclude that
$\omega$ restricts to a symplectic form on $\fk_{\bar 1}$ and hence
$\fk_{\bar 1}$ generates a maximal superalgebra isomorphic to
$\fosp(1|16)$.


\subsection{F1-string and NS5-brane, Weyl transformations and near-horizon limits}
\label{sec:10dCSGoffshellHalfBPSSUSY} 

Backgrounds which solve \eqref{eq:10dCSGvacua} for precisely eight
linearly independent supersymmetry parameters $\epsilon$ are called
half-BPS. Two well-known half-BPS backgrounds in ten dimensions are
the F1-string \cite{Dabholkar:1990yf} and the NS5-brane
\cite{Duff:1990wv}. They solve \eqref{eq:10dCSGvacua} with $\rd H =0$
and $G \epsilon = \half H \epsilon$ and thus define half-BPS
backgrounds of type I supergravity in ten dimensions.  To define them,
it is convenient to write $g_{\RR^{p,q}}$ and $\vol_{\RR^{p,q}}$ for
the canonical flat metric and volume form on $\RR^{p,q}$.

The F1-string background has metric and three-form given by
\begin{equation}\label{eq:StringMetric}
  g = \te^{2\Phi} g_{\RR^{1,1}} + g_{\RR^{8}} \; , \quad\quad H =
  \vol_{\RR^{1,1}} \wedge \,\rd \te^{2\Phi}~,
\end{equation}
where $\te^{-2\Phi}$ is a harmonic function on $\RR^8$ so
that $\rd ( \te^{-2\Phi} \star H )=0$. For example, thinking of
$\RR^8$ as a cone over $S^7$ with radial coordinate $r$, one can take
$\te^{-2\Phi} = 1 + \tfrac{| k_2 |}{r^6}$ for some constant
$k_2$. The supersymmetry parameter is given by
\begin{equation}\label{eq:StringSUSY}
  \epsilon = \te^{\Phi/2} \epsilon_0 \; , \quad\quad \vol_{\RR^{1,1}}
  \epsilon_0 = \epsilon_0~,
\end{equation}
where $\epsilon_0$ is a constant positive chirality Majorana--Weyl
spinor on $\RR^{9,1}$.

Now consider the Weyl transformation (with $\Omega =
\te^{-\Phi/3}$) of the F1-string  that defines a solution of
\eqref{eq:10dCSGvacuaGzero}.  This is a new half-BPS background of
conformal (but not Poincaré) supergravity in ten dimensions. Its
`near-horizon' limit is defined by taking the radial coordinate
$r\rightarrow 0$, which recovers precisely the maximally
supersymmetric $\AdS_3 \times S^7$ background obtained in
Section~\ref{sec:10dCSGoffshellMaxSUSY} (identifying $|k_2|^{-1/3} =
R/18$).

The NS5-brane background has metric and three-form given by
\begin{equation}\label{eq:5braneMetric}
  g = g_{\RR^{5,1}} + \te^{2\Phi} g_{\RR^{4}} \; , \quad\quad H = -
  {\star _{\RR^4} \rd} \te^{2\Phi}~,
\end{equation}
where $\te^{2\Phi}$ is a harmonic function on $\RR^4$ so that
$\rd H=0$.  For example, thinking of $\RR^4$ as a cone over $S^3$ with
radial coordinate $r$, one can take $\te^{2\Phi} = 1 +
\tfrac{| k_6 |}{r^2}$ for some constant $k_6$. The supersymmetry
parameter is given by
\begin{equation}\label{eq:5braneSUSY}
\epsilon = \epsilon_0 \; , \quad\quad \vol_{\RR^{4}} \epsilon_0 = \epsilon_0~, 
\end{equation}
where $\epsilon_0$ is a constant positive chirality Majorana--Weyl
spinor on $\RR^{9,1}$.

The near-horizon limit of the NS5-brane defines a metric on $\RR^{5,1}
\times \RR_+ \times S^3$ and is therefore not conformally equivalent
to the maximally supersymmetric $\AdS_7 \times S^3$ background
obtained Section~\ref{sec:10dCSGoffshellMaxSUSY}. However, it is
important to stress that any choice of function $\te^{2\Phi}$
on $\RR^4$ for the NS5-brane defines a half-BPS background of
conformal supergravity in ten dimensions. Let us therefore not assume
that $\te^{2\Phi}$ is harmonic on $\RR^4$ and perform the
Weyl transformation (with $\Omega = \te^{-\Phi/3}$) to define
a solution of \eqref{eq:10dCSGvacuaGzero}. Now, for this new half-BPS
background of conformal supergravity, taking $\te^{2\Phi} = 1
+ \tfrac{| k^\prime_6 |}{r^3}$ for some constant $k^\prime_6$ (which
is not harmonic on $\RR^4$), one recovers in the near-horizon limit
precisely the maximally supersymmetric $\AdS_7 \times S^3$ background
obtained in Section~\ref{sec:10dCSGoffshellMaxSUSY} (identifying
$|k^\prime_6|^{-2/3} = -2R/9$).

\section{Yang-Mills supermultiplet}
\label{sec:10dSYMonshell} 

The on-shell Yang-Mills supermultiplet in ten dimensions contains a
bosonic gauge field $A_\mu$ and a fermionic Majorana--Weyl spinor
$\lambda$ (we take $\lambda$ with positive chirality, i.e., $\Gamma
\lambda = \lambda$). Both fields are valued in a real Lie algebra
$\fg$ with invariant inner product $(-,-)$.

The supersymmetry variations are     
\begin{equation}
  \label{eq:10dsusy}
  \begin{split}
    \delta_\epsilon A_\mu &= {\overline \epsilon} \Gamma_\mu \lambda\\
    \delta_\epsilon \lambda &= - F \epsilon~,
  \end{split}
\end{equation}
where $\epsilon$ is a bosonic Majorana--Weyl spinor with positive
chirality. The variations in \eqref{eq:10dsusy} are Weyl-invariant
provided $(A_\mu,\lambda,\epsilon)$ are assigned weights
$(0,-\tfrac{3}{2},\half)$. For a bosonic supersymmetric conformal
supergravity background, the supersymmetry parameter $\epsilon$ obeys
\eqref{eq:10dCSGvacua}.

Up to boundary terms, the lagrangian
\begin{equation}\label{eq:10dsusylagcurved}
L = \te^{-2\Phi} \left( -\tfrac{1}{4} ( F_{\mu\nu} ,
  F^{\mu\nu} ) -\half ( {\overline \lambda} , \DiracD \lambda ) +
  \tfrac{1}{8} ( {\overline \lambda} , H \lambda ) + \half
  H^{\mu\nu\rho} ( A_\mu , \partial_\nu A_\rho + \tfrac{1}{3} [ A_\nu
  , A_\rho ] ) \right)~,
\end{equation}
is invariant under \eqref{eq:10dsusy}, for any $\epsilon$ obeying
\eqref{eq:10dCSGvacua}. (This result was noted in
\cite{deMedeiros:2012sb} for the subclass of bosonic supersymmetric
backgrounds of type I supergravity in ten dimensions.) The prefactor
$\te^{-2\Phi}$ acts as an effective gauge coupling in
\eqref{eq:10dsusylagcurved}. For generic backgrounds with $H \neq 0$,
notice that rigid supersymmetry necessitates both a mass term for
$\lambda$ and a Chern-Simons coupling for the gauge field. Closure of
$\te^{-2\Phi} \star H$ ensures that the Chern-Simons coupling is
gauge-invariant.

Squaring $\delta_\epsilon$ in \eqref{eq:10dsusy} with $\epsilon$
subject to \eqref{eq:10dCSGvacua} gives
\begin{equation}
  \label{eq:10dsusysquared}
  \begin{split}
    \delta_\epsilon^2 A_\mu &= - F_{\mu\nu} \xi^\nu = \cL_\xi A_\mu + D_\mu \Lambda\\
    \delta_\epsilon^2 \lambda &= \cL_\xi \lambda + \half G_\xi \lambda
    + [ \lambda , \Lambda ] + ( \epsilon {\overline \epsilon} - \half
    \xi ) ( \DiracD \lambda - G \lambda - \tfrac{1}{4} H \lambda )~,
  \end{split}
\end{equation}
where $\xi^\mu = {\overline \epsilon} \Gamma^\mu \epsilon$ and
$\Lambda = - A_\xi$. The Lie derivative $\cL_X$ along a conformal
Killing vector $X$ is defined such that
$\cL_X A_\mu = X^\nu \partial_\nu A_\mu + ( \partial_\mu X^\nu )
A_\nu$ and $\cL_X \lambda = \nabla_X \lambda + \tfrac{1}{4} (
\nabla_\mu X_\nu ) \Gamma^{\mu\nu} \lambda$.  The final term on the
right hand side of $\delta_\epsilon^2 \lambda$ in
\eqref{eq:10dsusysquared} vanishes using the field equation $\DiracD
\lambda - G \lambda - \tfrac{1}{4} H \lambda =0$ for $\lambda$,
derived from \eqref{eq:10dsusylagcurved}. Thus, on-shell, it follows
that
\begin{equation} 
  \label{eq:10dsusyclosure}
  \delta_\epsilon^2 = \cL_\xi  + w \sigma_\xi  + \delta_\Lambda
\end{equation}
on any field in the Yang-Mills supermultiplet with Weyl weight $w$,
where $\sigma_\xi = - \tfrac{1}{10} \nabla_\mu \xi^\mu = -\tfrac{1}{3}
\partial_\xi \Phi$ is the parameter for a Weyl variation and
$\delta_\Lambda$ denotes a gauge variation with parameter $\Lambda =
-A_\xi$.

A novel (partially) off-shell formulation of supersymmetric Yang-Mills
theory on $\RR^{9,1}$ was obtained by Berkovits in
\cite{Berkovits:1993zz} (see also \cite{Evans:1994cb,Baulieu:2007ew}).
To match the $16$ off-shell fermionic degrees of freedom of $\lambda$,
the $9$ off-shell degrees of freedom of $A_\mu$ are supplemented by
$7$ bosonic auxiliary scalar fields $Y_i$ (where $i=1,\dots,7$). All
fields are $\fg$-valued. The supersymmetry parameter $\epsilon$ is
also supplemented by seven linearly independent bosonic Majorana--Weyl
spinors $\theta_i$, each with the same positive chirality as
$\epsilon$. The index $i$ corresponds to the vector representation of
the $\fspin(7)$ factor in the isotropy algebra $\fspin(7) \ltimes \RR^8$
of $\epsilon$.

Now consider the following supersymmetry variations for the partially
off-shell Yang-Mills supermultiplet on a bosonic supersymmetric
conformal supergravity background
\begin{equation}
  \label{eq:10dsusyoffshell}
  \begin{split}
    \delta_{\epsilon} A_\mu &= {\overline \epsilon} \Gamma_\mu \lambda\\
    \delta_{\epsilon} \lambda &= - F \epsilon + Y_i \theta_i \\
    \delta_{\epsilon} Y_i &= {\overline \theta_i} ( \DiracD \lambda -
    G \lambda - \tfrac{1}{4} H \lambda )~.
  \end{split}
\end{equation}

Under the Weyl transformation $g_{\mu\nu} \mapsto \Omega^2
g_{\mu\nu}$, $H_{\mu\nu\rho} \mapsto \Omega^2 H_{\mu\nu\rho}$, $\Phi
\mapsto \Phi + 3 \ln \Omega$ of the background data that was described
in Section~\ref{sec:10dCSGoffshell},  the supersymmetry variations in
\eqref{eq:10dsusyoffshell} are invariant provided we assign
$(A_\mu,\lambda,Y_i)$ their canonical weights $(0,-\tfrac{3}{2},-2)$,
with $\epsilon$ and $\theta_i$ both having weight $\half$. (The Weyl
transformation $\Dirac \mapsto \Omega^{-11/2} \Dirac \Omega^{9/2}$ of
the Dirac operator in ten dimensions can be used to prove this for
$\delta_\epsilon Y_i$.)

The supersymmetry parameters $\epsilon$ and $\theta_i$ are related
such that
\begin{equation}
  \label{eq:10doffshellnuidentity}
  {\overline \epsilon} \Gamma_\mu \theta_i = 0 \; , \qquad {\overline
    \theta_i} \Gamma_\mu \theta_j = \delta_{ij} \, \xi_\mu
  \qquad\text{and}\qquad \epsilon {\overline \epsilon} + \theta_i
  {\overline \theta_i} = \half \xi~.
\end{equation}

Squaring \eqref{eq:10dsusyoffshell} subject to
\eqref{eq:10doffshellnuidentity} gives precisely
\eqref{eq:10dsusyclosure} on $A_\mu$ and $\lambda$, without needing to
impose the field equation for $\lambda$. Moreover,
\begin{equation}
  \label{eq:10dsusyclosureY}
  \delta_\epsilon^2 Y_i = \cL_\xi Y_i -2 \sigma_\xi Y_i + [ Y_i , \Lambda ]  + \Upsilon_{ij} Y_j~,
\end{equation}
where $\Upsilon_{ij} = {\overline \theta}_{[i} \Dirac \theta_{j]} -
\tfrac{1}{4} {\overline \theta}_{i} H \theta_{j}$ corresponds to a
$\fspin(7)$ rotation.

Up to boundary terms, the lagrangian
\begin{multline}
  \label{eq:10dsusylagcurvedOffShell}
  L = \te^{-2\Phi} \left( -\tfrac{1}{4} ( F_{\mu\nu} , F^{\mu\nu} )
    -\half ( {\overline \lambda} , \DiracD \lambda ) + \half ( Y_i ,
    Y_i ) + \tfrac{1}{8} ( {\overline \lambda} , H \lambda ) \right.\\
    \left. {} + \half H^{\mu\nu\rho} ( A_\mu , \partial_\nu A_\rho +
    \tfrac{1}{3} [ A_\nu , A_\rho ] ) \right)~,
\end{multline}
is invariant under \eqref{eq:10dsusyoffshell}. It is also manifestly
invariant under $\fspin(7)$ rotations of the auxiliary fields.  Moreover,
the integral of \eqref{eq:10dsusylagcurvedOffShell} is Weyl-invariant
with respect to the aforementioned transformation rules for fields and
background data.

Of all the bosonic supersymmetric conformal supergravity backgrounds
the Yang--Mills supermultiplet above can be defined upon, the
maximally supersymmetric $\AdS_3 \times S^7$ and $\AdS_7 \times S^3$
Freund--Rubin backgrounds classified in
Section~\ref{sec:10dCSGoffshellMaxSUSY} are perhaps the most
compelling. In particular, it would interesting to explore whether the
Yang--Mills supermultiplet on these conformal supergravity backgrounds
admits a consistent truncation that would recover one of the theories
described in \cite{Blau:2000xg, Ito:2012hs, Fujitsuka:2012wg,
  Minahan:2015jta,Minahan:2015any}. The relevant theories in \cite{Blau:2000xg} (or
{\cite{Fujitsuka:2012wg}}) would follow by dimensionally reducing the
on-shell (or partially off-shell) Yang--Mills supermultiplet on
$\RR^{9,1}$ to some lower dimension $d$ equal to either $7$ or $3$,
before deforming the resulting supermultiplet in dimension $d$ in such
a way that it retains rigid supersymmetry on a curved space admitting
the maximum number of real or imaginary Killing spinors, i.e., either
$AdS_d$ or $S^d$. The deformation involves introducing several
non-minimal couplings that do not seem to figure in \eqref{eq:10dsusy}
and \eqref{eq:10dsusylagcurved}, though this discrepancy may be the
result of a non-standard reduction along some subset of Killing
vectors of $S^{10-d}$ or $AdS_{10-d}$ that is necessary for the
conformal supergravity background instead of along the obvious
translations in $\RR^{10-d}$ or $\RR^{9-d,1}$, as in
\cite{Blau:2000xg, Fujitsuka:2012wg}. We leave this question for
future work.


\section{Lifting to eleven dimensions}
\label{sec:11dLift} 

It should not have gone unnoticed that the supersymmetric backgrounds
of conformal supergravity in ten dimensions that we have been
discussing bear a striking resemblance to supersymmetric backgrounds
of Poincaré supergravity in eleven dimensions.  For instance, each
maximally supersymmetric background obtained in
Section~\ref{sec:10dCSGoffshellMaxSUSY} has an obvious maximally
supersymmetric counterpart in Theorem~1 of
\cite{FigueroaO'Farrill:2002ft}.  Moreover, the structure of the
half-BPS string and five-brane backgrounds obtained in
Section~\ref{sec:10dCSGoffshellHalfBPSSUSY} is virtually identical to
that of the well-known half-BPS M2-brane and M5-brane solutions of
Poincaré supergravity in eleven dimensions.  

This empirical evidence hints at an embedding of (at least some)
supersymmetric backgrounds of ten-dimensional conformal supergravity
in supersymmetric solutions of eleven-dimensional Poincaré
supergravity.  Of course, this would be distinct from the well-known
Kaluza--Klein reduction along a spacelike Killing vector for
supergravity backgrounds in eleven dimensions, yielding backgrounds of
type IIA supergravity in ten dimensions. After a brief synopsis of the
defining conditions for bosonic supersymmetric backgrounds and
solutions of eleven-dimensional Poincaré supergravity, we shall spend
the rest of this final section investigating a few different types of
embedding for some of the backgrounds of ten-dimensional conformal
supergravity that we have already encountered. This will begin with a
review of the Kaluza--Klein embedding of supersymmetric backgrounds of
type I supergravity. We will then describe a novel `equatorial'
embedding for the maximally supersymmetric Freund--Rubin backgrounds
of ten-dimensional conformal supergravity in their eleven-dimensional
counterparts. Finally, we will describe the embedding of the half-BPS
string and five-brane backgrounds of ten-dimensional conformal
supergravity and show how to recover the maximally supersymmetric
Freund--Rubin backgrounds via delocalisation and near-horizon limits.


\subsection{Supersymmetric solutions in eleven dimensions}
\label{sec:11dSUSYsolutions} 

The bosonic fields of Poincaré supergravity in eleven dimensions
consist of a metric ${\hat g}$ and a closed four-form ${\hat F}$.
Following the conventions of \cite{FigueroaO'Farrill:2002ft}, a
bosonic supersymmetric background is given by a solution of
\begin{equation}
  \label{eq:11dKS}
  {\hat \nabla}_M {\hat \epsilon} = -\tfrac{1}{24} {\hat{\Gamma}}_M
  {\hat F} {\hat \epsilon} + \tfrac{1}{8} {\hat F}  {\hat{\Gamma}}_M
  {\hat \epsilon}~,
\end{equation}
where ${\hat \epsilon}$ is a Majorana spinor in eleven dimensions. Any
such background is called a supersymmetric solution if it also obeys
the field equations
\begin{equation}
  \label{eq:11dEOM}
  \begin{split}
    {\hat R}_{MN} &= \tfrac{1}{12} {\hat F}_{MABC} {\hat F}_{N}{}^{ABC} - \tfrac{1}{144} {\hat g}_{MN} {\hat F}_{ABCD} {\hat F}^{ABCD}\\
    \rd {{\hat \star }{\hat F}} &= - \half {\hat F} \wedge {\hat F}~.
  \end{split}
\end{equation}
Note that both \eqref{eq:11dKS} and \eqref{eq:11dEOM} are invariant
under the homothety $({\hat g},{\hat F}) \mapsto (\alpha^2 {\hat g},
\alpha^3 {\hat F})$, for any $\alpha \in \RR^\times$.


\subsection{Kaluza--Klein embedding of supersymmetric type I backgrounds}
\label{sec:11dTypeIKKUplift} 

It is well-known that any supersymmetric solution of type IIA
supergravity in ten dimensions can be uplifted to a supersymmetric
solution of supergravity in eleven dimensions via the `string-frame'
Kaluza--Klein ansatz. This recovers only the subset of supersymmetric
solutions of supergravity in eleven dimensions which admit a spacelike
Killing vector $\xi$ with $\cL_\xi {\hat F} =0$. At least locally, one
can write $\xi = \partial_z$ in terms of the eleventh coordinate $z$.

Now consider the following special case of the aforementioned ansatz: 
\begin{equation}
  \label{eq:11dAnsatz}
  \begin{split}
    {\hat g} &= \te^{4\Phi/3} (\rd z)^2 + \te^{-2\Phi/3} g \\
    {\hat F} &= \rd z \wedge H~,
  \end{split}
\end{equation}
in terms of a metric $g$, a function $\Phi$ and a three-form $H$ in
ten dimensions. It follows that ${\hat \star } {\hat F} =
\te^{-2\Phi} {\star H}$. Plugging \eqref{eq:11dAnsatz} into the
second field equation in \eqref{eq:11dEOM} therefore gives $\rd (
\te^{-2\Phi}  {\star H} ) =0$. It also follows that $\rd H =0$
since ${\hat F}$ is closed.

The ansatz \eqref{eq:11dAnsatz} allows one to define an idempotent
element ${\bf I} = \te^{-2\Phi/3} {\hat \Gamma}_z$ which
anticommutes with every ${\hat \Gamma}_\mu$ (where $\mu$ is any index
$M \neq z$). If ${\hat \epsilon} = {\bf I} {\hat \epsilon}$ then it can be
identified with a positive chirality Majorana--Weyl spinor
$\te^{-\Phi/6} \epsilon$ in ten dimensions. Assuming this to
be the case then plugging \eqref{eq:11dAnsatz} into \eqref{eq:11dKS}
gives
\begin{equation}
  \label{eq:11dAnsatzKS}
  \begin{split}
    \nabla_\mu \epsilon &= \tfrac{1}{6} \Gamma_\mu G \epsilon +
    \tfrac{1}{24} \Gamma_\mu H \epsilon + \tfrac{1}{8} H \Gamma_\mu
    \epsilon \\
    G \epsilon &= \tfrac{1}{2} H \epsilon~,
  \end{split}
\end{equation}
with $\partial_z \epsilon =0$ and $G = \rd \Phi$. The first condition
in \eqref{eq:11dAnsatzKS} is identified with \eqref{eq:10dCSGvacua}
provided $\rd ( \te^{-2\Phi} {\star H} ) =0$. Since $H$ is closed,
the second condition in \eqref{eq:11dAnsatzKS} then gives precisely
the defining condition for a bosonic supersymmetric background of type
I supergravity in ten dimensions.

To summarise, we have shown that any bosonic supersymmetric background
of type I supergravity in ten dimensions can be embedded via
\eqref{eq:11dAnsatz} in a bosonic supersymmetric background of
Poincaré supergravity in eleven dimensions, obeying
\eqref{eq:11dKS} for some ${\hat \epsilon} = {\bf I} {\hat \epsilon}$ and
the second field equation in \eqref{eq:11dEOM}. Of course, the
projection condition in eleven dimensions is because any background of
type I supergravity in ten dimensions can preserve no more than
sixteen real supercharges (in contrast with the maximum of thirty two
in eleven dimensions).


\subsection{Embedding of maximally supersymmetric Freund--Rubin backgrounds}
\label{sec:11dMaxSUSYFR}

Poincaré supergravity in eleven dimensions admits two well-known
maximally supersymmetric Freund--Rubin solutions. In terms of the
scalar curvature ${\hat R}$ of ${\hat g}$, they are of the form
\begin{itemize}
\item $\AdS_4 ( 8{\hat R} ) \times S^7 (-7{\hat R} )$ with ${\hat F} = \sqrt{-6{\hat R}} \, \vol_{\AdS_4}$ (if ${\hat R}<0$).
\item $\AdS_7 ( -7{\hat R} ) \times S^4 (8{\hat R} )$ with ${\hat F} = \sqrt{6{\hat R}} \, \vol_{S^4}$ (if ${\hat R}>0$).
\end{itemize}
(The scalar curvature of each $\AdS$ and $S$ factor is denoted in
parenthesis.)

To make our description of the embedding as transparent as possible,
let us adopt the following notation. Let $g_n$ denote the `unit
radius' metric on either $\AdS_n$ or $S^n$ (i.e., the metric with
constant scalar curvature $-n(n-1)$ for $\AdS_n$ or $n(n-1)$ for
$S^n$). Any metric of the form $\kappa^2 g_n + \lambda^2 g_m$ will be
assumed to be Lorentzian (i.e., $\AdS_n \times S^m$ or $S^n \times
\AdS_m$). Let $\vol_n$ denote the volume form with respect to
$g_n$. Let $\psi_n$ denote a Killing spinor with respect to $g_n$,
obeying $\nabla_\mu \psi_n = \pm \half \Gamma_\mu \psi_n$ for $\AdS_n$
or $\nabla_\mu \psi_n = \pm \tfrac{i}{2} \Gamma_\mu \psi_n$ for
$S^n$. For $\AdS_n$ (or $S^n$), $\psi_n$ lifts to a constant spinor on
the flat cone $C(\AdS_n) \cong \RR^{n-1,2}$ (or $C(S^n) \cong
\RR^{n+1}$). We shall refer to $\psi_n$ as having unit Killing
constant. Rescaling $g_n$ by a factor of $\kappa^2$ rescales the
Killing constant by a factor of $\kappa^{-1}$.

In terms of this notation, the data for the maximally supersymmetric
Freund--Rubin solutions of eleven-dimensional Poincaré supergravity is
given by
\begin{equation}
  \label{eq:11dFRgF}
  {\hat g} = {\hat \kappa}^2 ( g_4 + 4 g_7 ) 
  \qquad\text{and}\qquad {\hat F} = 3 {\hat \kappa}^3 \vol_4~,
\end{equation}
while the supersymmetry parameter ${\hat \epsilon}$ involves a tensor
product of $\psi_4$ (with Killing constant ${\hat \kappa}^{-1}$) and
$\psi_7$ (with Killing constant $(2{\hat \kappa})^{-1}$). The constant
\begin{equation}
  \label{eq:11dkappa}
  {\hat \kappa} := \sqrt{\frac{3}{2|{\hat R}|}}~.
\end{equation}
Observe that the factors of ${\hat \kappa}$ above are precisely the
same as for the homothety noted at the end of
Section~\ref{sec:11dSUSYsolutions}, hence we can and will fix ${\hat
  \kappa} =1$ via the action of a homothety with $\alpha = {\hat 
  \kappa}^{-1}$.

On the other hand, the data for the maximally supersymmetric
Freund--Rubin backgrounds of ten-dimensional conformal supergravity is
given by
\begin{equation}
  \label{eq:10dFRgH}
  g = \kappa^2 ( g_3 + 4 g_7 ) 
  \qquad\text{and}\qquad H = 3 \kappa^2 \vol_3~,
\end{equation}
while the supersymmetry parameter $\epsilon$ involves a tensor product
of $\psi_3$ (with Killing constant $\kappa^{-1}$) and $\psi_7$ (with
Killing constant $(2\kappa)^{-1}$). The constant
\begin{equation}
  \label{eq:10dkappa}
  \kappa := \sqrt{\frac{9}{2|R|}}~.
\end{equation}
Since $G=0$ for this class of backgrounds, notice that the factors of
$\kappa$ above are precisely the same as for a (constant) Weyl
transformation. Therefore we shall fix $\kappa =1$ via a Weyl
transformation with $\Omega = \kappa^{-1}$.

To embed \eqref{eq:10dFRgH} (with $\kappa =1$) in \eqref{eq:11dFRgF}
(with ${\hat \kappa}=1$), it remains only to recognise the canonical
`equatorial' embedding defined by
\begin{equation}\label{eq:equatorialg}
  g_4 = \rd z^2 + f(z)^2 g_3~,
\end{equation}
where $f(z)$ is $\cosh(z)$ for $\AdS_3 \subset \AdS_4$ or $\cos(z)$
for $S^3 \subset S^4$, in terms of the `colatitude' $z$. From
\eqref{eq:equatorialg}, it follows that $\vol_4 = f(z)^3 \rd z \wedge
\vol_3$ and hence that at $z=0$, we have
\begin{equation}
  \label{eq:11dFRgFembedding}
  {\hat g} =  \rd z^2 + g \qquad\text{and}\qquad {\hat F} = \rd z
  \wedge H~.
\end{equation}

The embedding of the supersymmetry parameter $\epsilon$ in ${\hat
  \epsilon}$ is prescribed by the embedding of the unit Killing spinor
$\psi_3$ in $\psi_4$ (the other Killing spinor $\psi_7$ clearly just
goes along for the ride). Recall that $\psi_3$ and $\psi_4$ are
completely specified by constant spinors on their respective (flat)
cones. By definition, in terms of a radial coordinate $r$, the
relevant cone metric $g_{C_{n+1}}$ is either $- \rd r^2 +r^2 g_n$ for
$\AdS_n$ or $\rd r^2 +r^2 g_n$ for $S^n$. For $\AdS_3 \subset \AdS_4$,
it follows that
\begin{equation}
  \label{eq:AdS4Cone}
  g_{C_5} = - \rd r^2 +r^2 g_4 = - \rd r^2 +r^2 \rd z^2 + ( r \cosh(z)
  )^2 g_3 = \rd x^2 - \rd y^2 + y^2 g_3 = \rd x^2 + g_{C_4}~,
\end{equation}
where $x = r \sinh(z)$ and $y = r \cosh(z)$. The embedding
$\Delta^{(2,2)}_+ \subset \Delta^{(3,2)}$ of Killing spinors here is
therefore prescribed by restricting $\Delta^{(3,2)}$ to the $x=0$
hyperplane in $C_5 \cong \RR^{3,2}$. Similarly, for $S^3 \subset S^4$,
it follows that
\begin{equation}
  \label{eq:S4Cone}
  g_{C_5} = \rd r^2 +r^2 g_4 = \rd r^2 +r^2 \rd z^2 + ( r \cos(z) )^2
  g_3 = \rd x^2 + \rd y^2 + y^2 g_3 = \rd x^2 + g_{C_4}~,
\end{equation}
where $x = r \sin(z)$ and $y = r \cos(z)$. Therefore the embedding
$\Delta^{(4)}_+ \subset \Delta^{(5)}$ of Killing spinors here is
prescribed by restricting $\Delta^{(5)}$ to the $x=0$ hyperplane in
$C_5 \cong \RR^{5}$.


\subsection{Branes, delocalisation and near-horizon limits}
\label{sec:11dHalfBPSSUSY} 

Two well-known half-BPS solutions of supergravity in eleven dimensions
are the M2-brane and the M5-brane.

The M2-brane solution has metric and four-form given by
\begin{equation}\label{eq:M2Metric}
  {\hat g} = f^{-2/3} g_{\RR^{2,1}} + f^{1/3} g_{\RR^{8}}
  \qquad\text{and}\qquad {\hat F} = \vol_{\RR^{2,1}} \wedge \rd
  f^{-1}~,
\end{equation}
where $f$ is a harmonic function on $\RR^8$ so that the second field
equation in \eqref{eq:11dEOM} is satisfied (i.e., $\rd {{\hat \star
  }{\hat F}} =0$ since ${\hat F} \wedge {\hat F} = 0$ for
\eqref{eq:M2Metric}). The supersymmetry parameter is given by
\begin{equation}\label{eq:M2SUSY}
  {\hat \epsilon} = f^{-1/6} {\hat \epsilon}_0
  \qquad\text{with}\qquad  \vol_{\RR^{2,1}} {\hat \epsilon}_0 = {\hat
    \epsilon}_0~,
\end{equation}
where ${\hat \epsilon}_0$ is a constant Majorana spinor on
$\RR^{10,1}$.

By identifying $z$ with a spatial coordinate on $\RR^{2,1}$ and
$f=\te^{-2\Phi}$, one recognises that \eqref{eq:M2Metric} is
of the form \eqref{eq:11dAnsatz}. With respect to these
identifications, the data $(g,H)$ in \eqref{eq:11dAnsatz} gives
precisely the F1-string background of type I supergravity in ten
dimensions in \eqref{eq:StringMetric}. The conformally related data
$(\te^{-2\Phi/3} g,\te^{-2\Phi/3} H)$ in ten
dimensions gives precisely the half-BPS background of conformal
supergravity with maximally supersymmetric $\AdS_3 \times S^7$
near-horizon limit. On the other hand, the near-horizon limit of
\eqref{eq:M2Metric} in eleven dimensions with
$f = \te^{-2\Phi} = 1 + \tfrac{| k_2 |}{r^6}$ gives the
maximally supersymmetric solution
$\AdS_4 ( 8{\hat R} ) \times S^7 (-7{\hat R} )$ with
${\hat F} = \sqrt{-6{\hat R}} \, \vol_{\AdS_4}$ (after
identifying $|k_2|^{-1/3} = -{\hat R}/6$).

The M5-brane solution has metric and four-form given by
\begin{equation}
  \label{eq:M5Metric}
  {\hat g} = f^{-1/3} g_{\RR^{5,1}} + f^{2/3} g_{\RR^{5}}
  \qquad\text{and}\qquad {\hat \star }{\hat F} = \vol_{\RR^{5,1}}
  \wedge \rd f^{-1}~,
\end{equation}
where $f$ is a harmonic function on $\RR^5$ so that $\rd {\hat F}
=0$.  The supersymmetry parameter is given by
\begin{equation}
  \label{eq:M5SUSY}
  {\hat \epsilon} = f^{-1/12} {\hat \epsilon}_0
  \qquad\text{with}\qquad \vol_{\RR^{5,1}} {\hat \epsilon}_0 = {\hat
    \epsilon}_0~,
\end{equation}
where again ${\hat \epsilon}_0$ is a constant Majorana spinor on
$\RR^{10,1}$.

By identifying $z$ with a coordinate on $\RR^5$ and $f = \te^{2\Phi}$,
one recognises that \eqref{eq:M5Metric} is of the form
\eqref{eq:11dAnsatz}. However, $\partial_z$ is a Killing vector only
if $f$ is harmonic on the subspace $\RR^4 \subset \RR^5$ orthogonal to
the $z$-direction. Making this assumption is known as `delocalisation'
along the $z$-direction. With respect to these identifications, the
data $(g,H)$ in \eqref{eq:11dAnsatz} for the delocalised M5-brane
gives precisely the NS5-brane background of type I supergravity in
ten dimensions in \eqref{eq:5braneMetric}. The near-horizon limit of
the delocalised M5-brane \eqref{eq:M5Metric} in eleven dimensions with
$f = \te^{2\Phi} = 1 + \tfrac{| k_6 |}{r^2}$ defines a half-BPS
background that is conformally equivalent to
$\RR^{5,1} \times H^2 \times S^3$ (c.f.
$\RR^{5,1} \times \RR_+ \times S^3$ in the near-horizon limit of the
NS5-brane in ten dimensions). On the other hand, without
delocalisation, the near-horizon limit of \eqref{eq:M5Metric} with
$f = \te^{2\Phi} = 1 + \tfrac{| k^\prime_6 |}{r^3}$ gives the
maximally supersymmetric solution
$\AdS_7 ( -7{\hat R} ) \times S^4 (8{\hat R} )$ with
${\hat F} = \sqrt{6{\hat R}} \, \vol_{S^4}$ (after identifying the
constant $k^\prime_6$ such that $|k^\prime_6|^{-2/3} = 2{\hat
  R}/3$).
Of course, without delocalisation, the ansatz \eqref{eq:11dAnsatz}
cannot be used to reduce to ten dimensions. Even so, notice that the
data $(\te^{-2\Phi/3} g,\te^{-2\Phi/3} H)$ in ten dimensions obtained
by comparing \eqref{eq:11dAnsatz} with \eqref{eq:M5Metric} without
delocalisation gives precisely the half-BPS background of conformal
supergravity with maximally supersymmetric $\AdS_7 \times S^3$
near-horizon limit.

\section*{Acknowledgments}

The research of JMF is supported in part by the grant ST/J000329/1
``Particle Theory at the Tait Institute'' from the UK Science and
Technology Facilities Council, which also funded the visit of PdM to
Edinburgh which started this collaboration.  During the final stretch
of writing, JMF was a guest of the Grupo de Investigación ``Geometría
diferencial y convexa'' of the Universidad de Murcia, and he wishes to
thank Ángel Ferrández Izquierdo for the invitation, the hospitality
and for providing such a pleasant working atmosphere.

\appendix

\section{Clifford algebra and spinors in ten dimensions}
\label{sec:10dCliffordSpinors} 

In ten dimensions and in lorentzian signature, the `mostly plus' and
`mostly minus' inner products result in isomorphic Clifford
algebras.  Indeed, as real associative algebras, the Clifford algebra
$\Cl(9,1) \cong \Cl(1,9) \cong \Mat_{32}(\RR)$.  We shall work with
$\Cl(9,1)$ in this paper, which is defined by the relation
\begin{equation}
  \label{eq:10dgammaDefiningRelation}
  \Gamma_\mu \Gamma_\nu + \Gamma_\nu \Gamma_\mu = + 2 \eta_{\mu\nu} \1~,
\end{equation}
where $\eta_{\mu\nu} = \diag(-1,\underbrace{+1,\dots ,+1}_9)$ has mostly
plus signature.  In particular, $(\Gamma_0)^2 = -\1$.  It follows from
the above isomorphism that $\Cl(9,1)$ has a unique irreducible module
up to equivalence: let's call it $\Delta^{(9,1)}$.  As a representation of the
spin group $\Spin(9,1) \subset \Cl(9,1)$, $\Delta^{(9,1)}$ decomposes as a
direct sum of two irreducible spinor representations
$\Delta^{(9,1)} = \Delta_+^{(9,1)} \oplus \Delta_-^{(9,1)}$, where the subspaces
$\Delta_\pm^{(9,1)} \subset \Delta^{(9,1)}$ correspond to the $\pm 1$-eigenspaces of
the idempotent `chirality matrix' $\Gamma = - \Gamma_0 \Gamma_1
\dots \Gamma_9$, which is not central in $\Cl(9,1)$ but does commute
with $\Spin(9,1)$.  In physics parlance, elements of $\Delta^{(9,1)}$ are
known as Majorana spinors and elements of $\Delta_\pm^{(9,1)}$ are known as
($\pm$ chirality) Majorana--Weyl spinors.

There exists on $\Delta^{(9,1)}$ a unique (up to an overall scale) symplectic
form $\bC$, the so-called charge conjugation matrix, which obeys
\begin{equation}
  \bC( \Gamma_\mu \psi, \chi ) = - \bC(\psi, \Gamma_\mu\chi )~.
\end{equation}
It follows that $\bC$ is $\Spin(9,1)$-invariant and, in addition, that
the chirality matrix $\Gamma$ is skew-symmetric.  This means that
$\Delta_\pm^{(9,1)}$ are lagrangian subspaces, so $\bC$ pairs $\Delta_+^{(9,1)}$
nondegenerately with $\Delta_-^{(9,1)}$, thus providing an isomorphism
$\Delta_+^{(9,1)} \cong (\Delta_-^{(9,1)} )^*$ of $\Spin(9,1)$ representations.

The Clifford algebra $\Cl(9,1)$ inherits a filtration from the tensor
algebra and the associated graded algebra is the exterior algebra of
$\RR^{9,1}$.  A convenient (vector space) isomorphism is provided by
the skewsymmetric products of the gamma matrices:
\begin{equation}
  \label{eq:10dgammak}
  \Gamma_{\mu_1 \dots \mu_k} = \Gamma_{[\mu_1} \dots \Gamma_{\mu_k ]}
  = \tfrac{1}{k!} \sum_{\sigma \in S_k} (-1)^{|\sigma|} \Gamma_{\mu_{\sigma(1)}}
  \dots \Gamma_{\mu_{\sigma(k)}}~,
\end{equation}
for degree $k>0$ elements (i.e., unit-weight skewsymmetrisation of $k$
distinct degree-one basis elements) and the identity element $\1$ for
$k=0$.  These form a basis for $\Cl(9,1)$. 

Some useful identities which follow are
\begin{equation}
  \label{eq:10dgammaidentities}
  \begin{split}
    \Gamma^\alpha \Gamma_{\mu_1 \dots \mu_k} \Gamma_\alpha &= (-1)^k
    (10-2k)\Gamma_{\mu_1 \dots \mu_k}\\
\Gamma^{\alpha\beta} \Gamma_{\mu_1 \dots \mu_k} \Gamma_{\alpha\beta} &= (10-(10-2k)^2)
    \Gamma_{\mu_1 \dots \mu_k}~,
  \end{split}
\end{equation} 
and
\begin{equation}
  \label{eq:10dgammadual}
  \Gamma_{\mu_1 \dots \mu_k} \Gamma = \sigma_{k-1} \tfrac{1}{(10-k)!}
   \varepsilon_{\mu_1 \dots \mu_k \nu_{k+1} \dots \nu_{10}}
  \Gamma^{\nu_{k+1} \dots \nu_{10}}~,
\end{equation}
where $\sigma_k = (-1)^{\lfloor \tfrac{k+1}{2} \rfloor}$ (i.e.,
$\sigma_1 = \sigma_2 = -1$, $\sigma_{k+2} = -\sigma_k$) and $\varepsilon_{01\dots 9} =+1$.

Now let us write ${\overline \psi} \chi = \bC (\psi,\chi)$, for any
$\psi , \chi \in \Delta^{(9,1)}$. It follows that
\begin{equation}
  \label{eq:10dbilnears}
  {\overline \psi} \Gamma_{\mu_1 \dots \mu_k} \chi = - \sigma_k \,
  {\overline \chi} \Gamma_{\mu_1 \dots \mu_k} \psi~,
\end{equation}
and
\begin{equation}
  \label{eq:10devenbilnears}
  {\overline \psi_\pm} \Gamma_{\mu_1 \dots \mu_{2k}} \chi_\pm = 0~,
\end{equation} 
for any $\psi_\pm , \chi_\pm \in \Delta_\pm^{(9,1)}$. Furthermore, we have the
Fierz identities
\begin{equation}
  \label{eq:10dfierz}
  \begin{split}
    \psi_\pm \, {\overline \chi_\pm} &= \tfrac{1}{32} \left( 2 (
      {\overline \chi_\pm} \Gamma^\mu \psi_\pm ) \Gamma_\mu -
      \tfrac{1}{3} ( {\overline \chi_\pm} \Gamma^{\mu\nu\rho} \psi_\pm
      ) \Gamma_{\mu\nu\rho} + \tfrac{1}{5!} ( {\overline \chi_\pm}
      \Gamma^{\mu\nu\rho\sigma\tau} \psi_\pm )
      \Gamma_{\mu\nu\rho\sigma\tau} \right) \bP_\mp\\
    \psi_\pm \, {\overline \chi_\mp} &= \tfrac{1}{16} \left( (
      {\overline \chi_\mp} \psi_\pm ) \1 - \tfrac{1}{2} (
      {\overline \chi_\mp} \Gamma^{\mu\nu} \psi_\pm ) \Gamma_{\mu\nu}
      + \tfrac{1}{4!} ( {\overline \chi_\mp} \Gamma^{\mu\nu\rho\sigma}
      \psi_\pm ) \Gamma_{\mu\nu\rho\sigma} \right) \bP_\pm~,
  \end{split}
\end{equation}
where $\bP_\pm = \half ( \1 \pm \Gamma )$. The bilinear
${\overline \chi_\pm} \Gamma_{\mu\nu\rho\sigma\tau} \psi_\pm$ defines
a five-form that is self-dual if the spinors have positive chirality
and anti-self-dual if the spinors have negative chirality.

It follows from \eqref{eq:10dbilnears} and
\eqref{eq:10devenbilnears} that, if $\epsilon \in \Delta_+^{(9,1)}$,
all bilinears built from $\epsilon$ vanish identically except for
$\xi_\mu = {\overline \epsilon} \Gamma_\mu \epsilon$ and
$\zeta_{\mu\nu\rho\sigma\tau} = {\overline \epsilon}
\Gamma_{\mu\nu\rho\sigma\tau} \epsilon$, which are nonzero for nonzero
$\epsilon$.  The Fierz identity~\eqref{eq:10dfierz} for $\epsilon$
reads
\begin{equation}
  \label{eq:10dfierzepsilon}
  \epsilon {\overline \epsilon} = \tfrac{1}{32} \left( 2 \xi + \zeta \right) \bP_-~,
\end{equation}
and a useful subsidiary identity is
\begin{equation}
  \label{eq:10dfierzepsilon2}
  - \Gamma^{\mu\nu} \epsilon {\overline \epsilon} \, \Gamma_\nu =
  \left( \left(  \epsilon {\overline \epsilon} - \half \xi
    \right) \Gamma^\mu + \xi^\mu \1 \right) \bP_+~.
\end{equation}

An unrelated source of $\pm$ signs comes from choosing a Witt (or
`lightcone') basis for $\RR^{9,1}$.  We choose a somewhat asymmetrical
definition:
\begin{equation}
  \Gamma_+ := \tfrac12 (\Gamma_9 + \Gamma_0) \qquad\text{and}\qquad
  \Gamma_- := \Gamma_9 - \Gamma_0~.
\end{equation}
It follows that $\Gamma_\pm^2 = 0$ and that
\begin{equation}
  \Gamma_+ \Gamma_- + \Gamma_- \Gamma_+ = 2 \1~.
\end{equation}
This last identity means that we may decompose $\Delta^{(9,1)}_+$ into the direct
sum of the two subspaces $\ker \Gamma_\pm : \Delta^{(9,1)}_+ \to
\Delta^{(9,1)}_-$, with $\tfrac12 \Gamma_\pm\Gamma_\mp$ the
corresponding projectors.


\providecommand{\href}[2]{#2}\begingroup\raggedright\endgroup

\end{document}